\documentclass[a4paper,final]{article}

\usepackage{graphicx}
\usepackage{amsmath}
\usepackage{amsgen}
\usepackage{amssymb}
\usepackage{amsfonts}
\usepackage{amsbsy}
\usepackage{subfigure}
\usepackage{exscale}
\usepackage{a4wide}
\usepackage[numbers]{natbib}

\DeclareMathAlphabet{\bi}{OML}{cmm}{b}{it}
\newfont{\tensy}{cmsy10}
\newcommand{\chem}[1]{{$\fontdimen16\tensy=3.0pt
    \fontdimen17\tensy=3.0pt \mathrm{#1}$}}

\renewcommand{\Im}[0]{\mathrm{Im}\,}
\newcommand{\etal}{{\rm et al.}\ }

\newcommand{\Tc}[0]{T_{\mathrm{C}}}
\newcommand{\up}[0]{\uparrow}
\newcommand{\Nup}[0]{\downarrow}
\newcommand{\en}[0]{\epsilon}
\newcommand{\om}[0]{\omega}
\newcommand{\Si}[0]{\Sigma}
\newcommand{\si}[0]{\sigma}
\newcommand{\las}[0]{\langle}
\newcommand{\ras}[0]{\rangle}
\newcommand{\la}[0]{\left\las}

\newcommand{\ra}[0]{\right\ras}

\newcommand{\Tr}[0]{\mathrm{Tr}}
\newcommand{\rme}{\mathrm{e}}
\newcommand{\rmi}{\mathrm{i}}
\newcommand{\rmd}{\mathrm{d}}

\newcommand{\bsigma}{\boldsymbol{\sigma}}

\begin{document}


\title{Ferromagnetism and Transport in the Double-Exchange Model,
  with and without Phonons; Application to the Manganites}
\author{D M Edwards and A C M Green\\Department of
    Mathematics, Imperial College, London SW7 2BZ, UK}
\maketitle
\begin{abstract}
  An introduction is given to the many-body coherent potential approximation
  (CPA) for the double-exchange (DE) model and the Holstein-DE model, the
  latter including coupling of the electrons to local phonons as well as to
  the local spins. It is shown how the method can treat the local spins and
  phonons quantum-mechanically and how it is equivalent to dynamical mean
  field theory in the classical limit. In the Holstein-DE model a full
  discussion is given of the cross-over from weak electron-phonon coupling
  through intermediate coupling, where small-polaron bands begin to appear,
  to strong coupling where some results similar to those of standard
  small-polaron theory are recovered. The theory is applied to ferromagnetic
  manganites with a full discussion of magnetic, transport, and spectroscopic
  data. It is found that many manganites are in the critical regime on the
  verge of small-polaron formation, which explains their sensitivity to
  parameters such as applied magnetic field and pressure. 
\end{abstract}

\tableofcontents

\newpage


\section{Introduction}

Magnetic materials exhibiting metallic behaviour can often be considered as
systems of local moments coupled to electrons in a conduction band by local
exchange interactions. The Hamiltonian for such a system is
\begin{equation}\label{eq:h_de}
  H_{\rm DE} = \sum_{ij\si} t_{ij}c_{i\si}^{\dag}c_{j\si}-J\sum_i
  \bi{S}_i\cdot\bsigma_i-h\sum_i\left(S_i^z+\si_i^z\right)\,,
\end{equation}
where $c^{\dag}_{i\si}$ creates an electron of spin $\si$ on lattice site
$i$, $\bi{S}_i$ is a local spin operator and
$\bsigma_i=\left(\si_i^x,\si_i^y,\si_i^z\right)$ is a conduction electron
spin operator defined by
\begin{equation}
  \si_i^+=\si_i^x+i\si_i^y=c_{i\up}^\dag
  c_{i\Nup}\,,\,\si_i^-=\si_i^x-i\si_i^y=c_{i\Nup}^\dag
  c_{i\up}\,,\,\si_i^z=\frac{1}{2} \left(n_{i\up}-n_{i\Nup}\right)
\end{equation}
with $n_{i\si}=c^\dag_{i\si}c_{i\si}$. The three terms of
equation~(\ref{eq:h_de}) describe hopping of the conduction electrons,
exchange coupling between local and itinerant spins and coupling to an
external magnetic field. If the local exchange coupling arises from
hybridization between the localized and itinerant electrons, as in anomalous
rare earth systems exhibiting heavy fermion behaviour, the exchange parameter
$J$ is negative. The Hamiltonian (\ref{eq:h_de}) is then often called the
Kondo lattice model in view of its connection with the Kondo impurity model
which has a local spin on one site only \cite{He93}. When Hund's rule
coupling is dominant $J>0$ and the system is sometimes called a ferromagnetic
Kondo lattice. This is misleading since for $J>0$ there is no connection with
the Kondo effect.

For $J>0$ it is useful to distinguish two distinct physical regimes,
depending on the magnitude of $J$ compared with the width $2W$ of the
conduction band. If $J\ll W$, as in a normal rare earth metal, the exchange
coupling can be treated as a perturbation which gives rise to the
Ruderman-Kittel-Kasuya-Yosida (RKKY) interaction between local moments. In
most rare earth metals this interaction, which oscillates in space, leads to
oscillatory or spiral configurations of the localized f electron moments.
The uniform ferromagnet \chem{Gd} is an exception. In this weak coupling
regime the Hamiltonian (\ref{eq:h_de}) is usually referred to as the $s-f$ or
$s-d$ model.

If $J\gg W$ the exchange coupling can no longer be treated as a perturbation.
A conduction electron can only hop onto a site with its spin parallel to the
local moment at that site. Furthermore if the number of conduction electrons
per atom $n\le1$ double occupation of a site is strongly suppressed. A single
electron at a site, with its spin parallel to the local spin $\bi{S}$, enjoys
an exchange energy $-JS/2$ which is lost if a second electron hops on.  The
system is therefore a {\em strongly correlated electron system}, just like
the Hubbard model in the regime of strong on-site Coulomb interaction $U$,
and for $n=1$ the system is a Mott insulator. In much of the theoretical work
on the present model the local spins are treated as classical vectors,
corresponding to $S\rightarrow\infty$. Since for $J\gg W$ the itinerant spin
must always be parallel to the local spin on each site, the effective hopping
integral for hopping between sites $i$ and $j$ becomes
$t_{ij}\cos\left(\theta_{ij}/2\right)$, where $\theta_{ij}$ is the angle
between the classical spins $\bi{S}_i$, $\bi{S}_j$. The cosine factor arises
from the scalar product of two spin $1/2$ eigenstates with different axes of
quantization. The resultant band narrowing in the paramagnetic state favours
ferromagnetism in order to lower the kinetic energy. This mechanism for
ferromagnetism was first introduced by Zener \cite{Ze51} and developed by
others \cite{AnHa55,Ge60,KuOh72}. Since it involves strong exchange coupling
on two adjacent atoms it is known as double-exchange. Consequently the
Hamiltonian (\ref{eq:h_de}) in the strong-coupling regime $J\gg W$ is called
the double-exchange (DE) model.  This paper is concerned with the DE model,
with quantum and classical local spins, and with its extension to include
coupling of the electrons to local phonons. We call this extended model the
Holstein-DE model.

The paper provides a simple introduction to the many-body coherent potential
approximation (CPA) described in detail in our previous papers
\cite{EdGrKu99,GrEd99,GrEd01,Gr01} and reviews the results. When spins and
phonons are treated classically the CPA is equivalent to the dynamical mean
field theory (DMFT) approach of Furukawa \cite{Fu94,Fu96} and Millis \etal
\cite{MiMuSh96II}. A fully quantum mechanical treatment within DMFT is a
difficult numerical task which has not yet been done. The many-body CPA may
be regarded as a useful analytic approximation to DMFT which becomes exact in
the classical limit.

The main application we have in mind is to the manganites \chem{La_{1-x}D_x
  MnO_3}, where \chem{D} is a divalent ion such as \chem{Ca} or \chem{Sr}.
Recently there has been renewed interest in these systems due to the
discovery of colossal magnetoresistance (CMR), the name given to a large
reduction in resistivity in an applied magnetic field \cite{Ra97}. For $x=0$,
\chem{LaMnO_3} is an antiferromagnetic insulator, but on doping with
$0.2<x<0.5$, the system becomes a ferromagnetic metal. The CMR phenomenon is
observed in the vicinity of the Curie temperature $\Tc$. In applying equation
(\ref{eq:h_de}) to these systems the local spins $\bi{S}_i$ are of magnitude
$S=3/2$, corresponding to three localized \chem{Mn} d electrons of $t_{2g}$
symmetry, and the band is derived from \chem{Mn} d states of $e_{g}$
symmetry. The band contains $n=1-x$ itinerant electrons per atom.
Since we are interested in doped ferromagnetic systems, the antiferromagnetic
interaction between neighbouring local spins plays no important role and is
neglected. A more serious simplification is the use of a single s band,
instead of two d bands based on orbitals of $e_{ g}$ symmetry. We discuss
this in section \ref{sec:appl-mang}.

In section \ref{sec:cpa-hubbard-model} we introduce the many-body CPA by
means of the Hubbard model and in section \ref{sec:many-body-cpa} we develop
it for the more complicated case of the DE model with quantum local spins.
In section \ref{sec:resist-param-state} we discuss the electrical resistivity
of the paramagnetic state in the DE model and show, as first stressed by
Millis \etal \cite{MiLiSh95}, that the DE model is unable to describe the
physics of the manganites completely. In section \ref{sec:magnetism-de-model}
we briefly discuss the variation of $\Tc$ as a function of band-filling $n$.
The development of the many-body CPA for the Holstein-DE model is discussed
in section \ref{sec:many-body-cpa-hde} and a comparison with standard
small-polaron theory is made in the limit of strong electron-phonon coupling.
For intermediate coupling we discuss the crossover, with increasing
temperature, from polaronic behaviour to a situation where the phonons behave
classically, the case considered by Millis \etal \cite{MiMuSh96II}. In
section \ref{sec:appl-mang} we consider the application of the Holstein-DE
model to the manganites. Comparison with experimental data leads to the
conclusion that typical manganites lie in the critical intermediate coupling
regime which is not fully described by previous theories. We summarise
briefly in section \ref{sec:conclusion}.


\section{CPA for the Hubbard model}\label{sec:cpa-hubbard-model}

To introduce the many-body CPA we consider the Hubbard model, which is a
simpler model for strongly correlated electrons than the DE model with
quantum spins. The Hamiltonian for this model is
\begin{equation}\label{eq:h_hubbard}
  H_{\rm H}=\sum_{ij\si}t_{ij}c^\dag_{i\si}c_{j\si}+U\sum_i n_{i\up}
  n_{i\Nup}
\end{equation}
and Hubbard \cite{Hu64} set out to find an approximation to the one-electron
retarded Green function $G_{\bi{k}\si}(\en)$ which is exact in the atomic
limit $t_{ij}=0$. He used the equation of motion method and the idea of
the alloy analogy described below. It turns out that Hubbard's approach,
without the minor ``resonance broadening correction'', is equivalent to the
CPA which was developed later \cite{ElKrLe74}. The CPA derivation of
Hubbard's result is much simpler than the original equation of motion
method. However we had to resort to an extension of the original method to
derive the many-body CPA for the DE model, with and without phonons, in the
general form needed to discuss magnetic properties. In this paper we restrict
the derivation to the paramagnetic state in zero magnetic field, although we
summarize some more general results.

The alloy analogy consists in considering the $\up$-spin electrons, say, to
move in the potential due to static $\Nup$-spin electrons, frozen in a random
configuration which must be averaged over. Thus a one-electron Hamiltonian
for $\up$-spin is obtained from (\ref{eq:h_hubbard}) by taking the last term
to be $U\sum_i n_{i\up} \las n_{i\Nup}\ras$ where $\las n_{i\Nup}\ras$ takes
the value $1$ with probability $n_{\Nup}$ and 0 with probability $1-n_\Nup$.
Here $n_\si$ is the number of $\si$-spin electrons per atom. It is important
to note that the alloy analogy is quite distinct from the Hartree-Fock
approximation in which $\las n_{i\si}\ras=n_\si$ for all $i$. In the alloy
analogy a $\si$-spin electron moves in a random potential given by $U$ on
$n_{\bar{\si}}$ sites and $0$ on $1-n_{\bar{\si}}$ sites. In the CPA the
random potential is replaced by a uniform, but energy-dependent, effective
potential $\Si_\si(\en)$ for an "effective medium". This effective potential,
in general complex, is called the coherent potential and is in fact the
electron self-energy. The procedure for determining $\Si_\si$ is to insist on
a zero average t-matrix for scattering by a central atom, with potential $U$
or $0$, set in the effective medium. Equivalently the average of the
site-diagonal element $G_\si(\en)$ of the Green function, for each type of
central atom, is put equal to the site-diagonal element of the Green function
for the effective medium. Thus
\begin{equation}\label{eq:cpa1}
  G_\si=n_{\bar{\si}}\frac{G_\si}{1-\left(U-\Si_\si\right)G_\si}+
  \left(1-n_{\bar{\si}}\right)\frac{G_\si}{1+\Si_\si G_\si}
\end{equation}
and
\begin{equation}\label{eq:cpa2}
  G_\si(\en)=\frac{1}{N}\sum_\bi{k}\frac{1}{\en-\en_\bi{k}-\Si_\si(\en)}=
  G_0\left(\en-\Si_\si(\en)\right)
\end{equation}
where the bare band Green function is given by
\begin{equation}\label{eq:cpa3}
  G_0(\en)=\int\,\rmd\en'\frac{N_0(\en')}{\en-\en'}\,.
\end{equation}
Here $\en_\bi{k}=\sum_{j}t_{ij}\exp[\rmi\bi{k}\cdot(\bi{R}_i-\bi{R}_j)]$ is
the band energy, where $\bi{R}_i$ is the position of site $i$, $N_0(\en)$ is
the corresponding density of states per atom and $N$ is the number of lattice
sites. Equations (\ref{eq:cpa1}) and (\ref{eq:cpa2}) are to be solved
self-consistently for $\Si_\si(\en)$ and hence for the local Green function
$G_\si(\en)$. Equation~(\ref{eq:cpa1}) may be written as
\begin{equation}\label{eq:cpa4}
  G_\si=\frac{n_{\bar{\si}}}{\Si_\si+G_\si^{-1}-U}+\frac{1-n_{\bar{\si}}}
  {\Si_\si+G_\si^{-1}}\,.
\end{equation}
This may be compared with the exact Green function for the atomic limit
($t_{ij}=0$) which is given by \cite{Hu64}
\begin{equation}\label{eq:cpa_AL}
  G_\si^{\rm AL}(\en)=\frac{n_{\bar{\si}}}{\en-U}+\frac{1-n_{\bar{\si}}}{\en}\,,
\end{equation}
where in this retarded Green function $\en$ has a small positive imaginary
part.  Hence
\begin{equation}\label{eq:cpa5}
  G_\si(\en)=G_\si^{\rm AL}\left(\Si_\si+G_\si^{-1}\right)\,.
\end{equation}
Clearly this CPA equation is exact in the atomic limit, when
$N_0(\en)=\delta(\en)$ and it follows from equation~(\ref{eq:cpa2}) that
$\Si_\si+G_\si^{-1}=\en$. Solution of the CPA equation becomes simple if the
density of states $N_0(\en)$ is taken to be of the elliptic form
\begin{equation}\label{eq:ell_dos}
  N_0(\en)=\frac{2}{\pi W^2}\left(W^2-\en^2\right)^{1/2}
\end{equation}
where $2W$ is the bandwidth. Then from equation~(\ref{eq:cpa3})
\begin{equation}\label{eq:G_0}
  G_0(\en)=\frac{2}{W^2}\left[\en-\left(\en^2-W^2\right)^{1/2}\right]\,.
\end{equation}
Introducing this expression for $G_\si$ in equation~(\ref{eq:cpa2}), and
solving for $\en-\Si_\si(\en)$, we find
\begin{equation}\label{eq:cpa6}
  \Si_\si+G_\si^{-1}=\en-W^2 G_\si/4
\end{equation}
Hence equations (\ref{eq:cpa_AL}) and (\ref{eq:cpa5}) give an algebraic
equation for $G_\si$.

Solving this type of equation for $G_\si$, Hubbard \cite{Hu64} calculated the
density of states $N_\si(\en)=-\pi^{-1}\Im G_\si(\en)$, considering
particularly the paramagnetic state $n_\up=n_\Nup=n/2$ and concentrating on
the half-filled band case $n=1$. He showed that for $U/W$ greater than a
critical value, equal to $1$ in the present approximation, a gap opens in
$N(\en)$ at the Fermi level so that the system becomes an insulator as
envisaged by Mott. For $U\gg W$ the density of states consists of two peaks
centred on $\en=0$ and $\en=U$, these being broadened versions of the
$\delta$-functions at these energies in the atomic limit. Furthermore, for
general band-filling $n$, the spectral weights in the two peaks are the same
as in the atomic limit.

The CPA for the Hubbard model has some serious defects. There are no
self-consistent solutions with magnetic order. Furthermore in the
paramagnetic metallic state, for $n<1$ or for $n=1$ with $U/W$ less than the
critical value, the system is never a Fermi liquid. There is never a sharp
Fermi surface at $T=0$ with a Migdal discontinuity in the Bloch state
occupation number, as pointed out by Edwards and Hewson \cite{EdHe68}. This
is due to the absence of states with infinite lifetime at the Fermi level,
since within the alloy analogy all states are scattered by disorder. A
modification of the CPA to remedy this defect, retaining the analytic
simplicity of the method, had some limited success
\cite{EdHe90,LuEd95,PoHeNo98}. However the most satisfactory approach is DMFT
which involves numerical solution of an associated self-consistent impurity
problem \cite{GeKoKrRo96,BuHePr98}. DMFT may be regarded as the best local
approximation, in which the self-energy is a function of energy only, and is
exact in infinite dimension.

The many-body CPA is considerably more satisfactory for the DE model than it
is for the Hubbard model, as discussed in the next session. There is one
limit, the case of classical spins ($S=\infty$), in which the CPA is
identical to DMFT. This is because classical spins are static and an alloy
analogy of frozen disordered spins is completely justified. DMFT for the DE
model has only been implemented for classical spins \cite{Fu94,Fu96} and the
many-body CPA discussed in the next section provides an approximate analytic
extension of DMFT to quantum spins. The system orders ferromagnetically below
a Curie temperature $\Tc$, as it should, and the disordered spin state above
$\Tc$ should be well described. However, the accuracy of the ground state at
$T=0$ for finite $S$ is unclear. The saturated state with all itinerant and
local spins completely aligned, which is the ground state for $S=\infty$ (we
are always considering large $J$ in the DE model), is never a self-consistent
CPA solution for finite $S$ \cite{EdGrKu99}. Actually the parameter range of
stability of the saturated ground state is unknown. It has been shown
rigorously that for $J=\infty$ it is unstable for $S=1/2$ and $0.12<n<0.45$,
with a simple cubic nearest-neighbour tight-binding band \cite{BrEd98}.  If
the true ground state is not saturated it seems unlikely to be a uniform
(spatially homogeneous) ferromagnet, with partially ordered local and
itinerant spins, as in the uniform CPA ground state for finite $S$. Such a
state would probably not be a Fermi liquid, just as in CPA, unless the
electrons making up the spin $S$ became partially delocalised with spectral
weight at the Fermi level. Such speculation goes beyond the DE model.


\section{The many-body CPA for the DE model}\label{sec:many-body-cpa}

In an earlier paper \cite{EdGrKu99} we developed the many-body CPA for the DE
model using an extension of Hubbard's equation~of motion method.
Hubbard's ``scattering correction'' becomes more complicated owing to the
form of the interaction term in the DE model whereby electrons can flip their
spin via exchange of angular momentum with the local spins. This dynamical
effect couples the equations for $G_\up$ and $G_\Nup$ and was first treated
by Kubo \cite{Ku74} in a one-electron dynamical CPA. The main feature of our
many-body CPA is that we recover Kubo's one-electron CPA as $n\rightarrow 0$
and the correct atomic limit for general band-filling $n$ as
$t_{ij}\rightarrow 0$. In a second paper \cite{GrEd99} we showed the
equivalence to DMFT in the limit $S\rightarrow\infty$, $J\rightarrow\infty$.
The full equation of motion derivation of the many-body CPA is required
to obtain general results in the presence of a magnetic field and/or magnetic
order \cite{GrEd99}. However it turns out that in the zero-field paramagnetic
state we can deduce the CPA equation from the atomic limit Green function
$G_\si^{\rm AL}$ and equation~(\ref{eq:cpa5}), just as in the Hubbard model.
We shall therefore not repeat the full derivation in this paper although we
shall discuss results on magnetic properties in
section \ref{sec:magnetism-de-model}.

The atomic limit Green function, $G_\up^{\rm AL}$, say, is easily obtained by
the equation of motion method using the Hamiltonian (\ref{eq:h_de}) with
$t_{ij}=0$ \cite{EdGrKu99}. The result for zero field ($h=0$) is
\begin{equation}\label{eq:de_AL_general}
  \begin{split}
    G_\up^{\rm AL}(\en)=\frac{1}{2S+1}\left[\frac{\la\left(S+S^z\right)
    n_\Nup-S^-\si^+\ra}{\en+J(S+1)/2}+\frac{\la\left(S-S^z\right)
    \left(1-n_\Nup\right)-S^-\si^+\ra}{\en-J(S+1)/2}\right.\\
    \left.{}+\frac{\la\left(S+1+S^z\right)\left(1-n_\Nup\right)+S^-\si^+\ra}
    {\en+JS/2}+\frac{\la\left(S+1-S^z\right)n_\Nup+S^-\si^+\ra}{\en-JS/2}\right]
  \end{split}  
\end{equation}
and for $h\ne 0$ one merely has to replace $\en$ by $\en+h/2$. The angle
brackets $\las\dots\ras$ represent thermal averages and all operators within
them correspond to the same site $i$, this suffix thus being omitted. This
expression, with four poles, is considerably more complicated than the
two-pole Hubbard model expression of equation~(\ref{eq:cpa_AL}). The poles at
$\en=\pm JS/2$, $\pm J(S+1)/2$ correspond to energies to add or remove an
electron from the atom, that is to transitions between singly-occupied states
and either unoccupied or doubly-occupied states. The singly-occupied states
have total spin $S+\frac{1}{2}$ or $S-\frac{1}{2}$ with energies $-JS/2$ and
$J(S+1)/2$ respectively; the unoccupied and doubly-occupied states have zero
energy.

\begin{figure}[htbp]
  \subfigure[]{\label{fig:DOS_DE_model_a}
  \begin{minipage}[b]{0.5\textwidth}
    \centering \includegraphics[width=\textwidth]{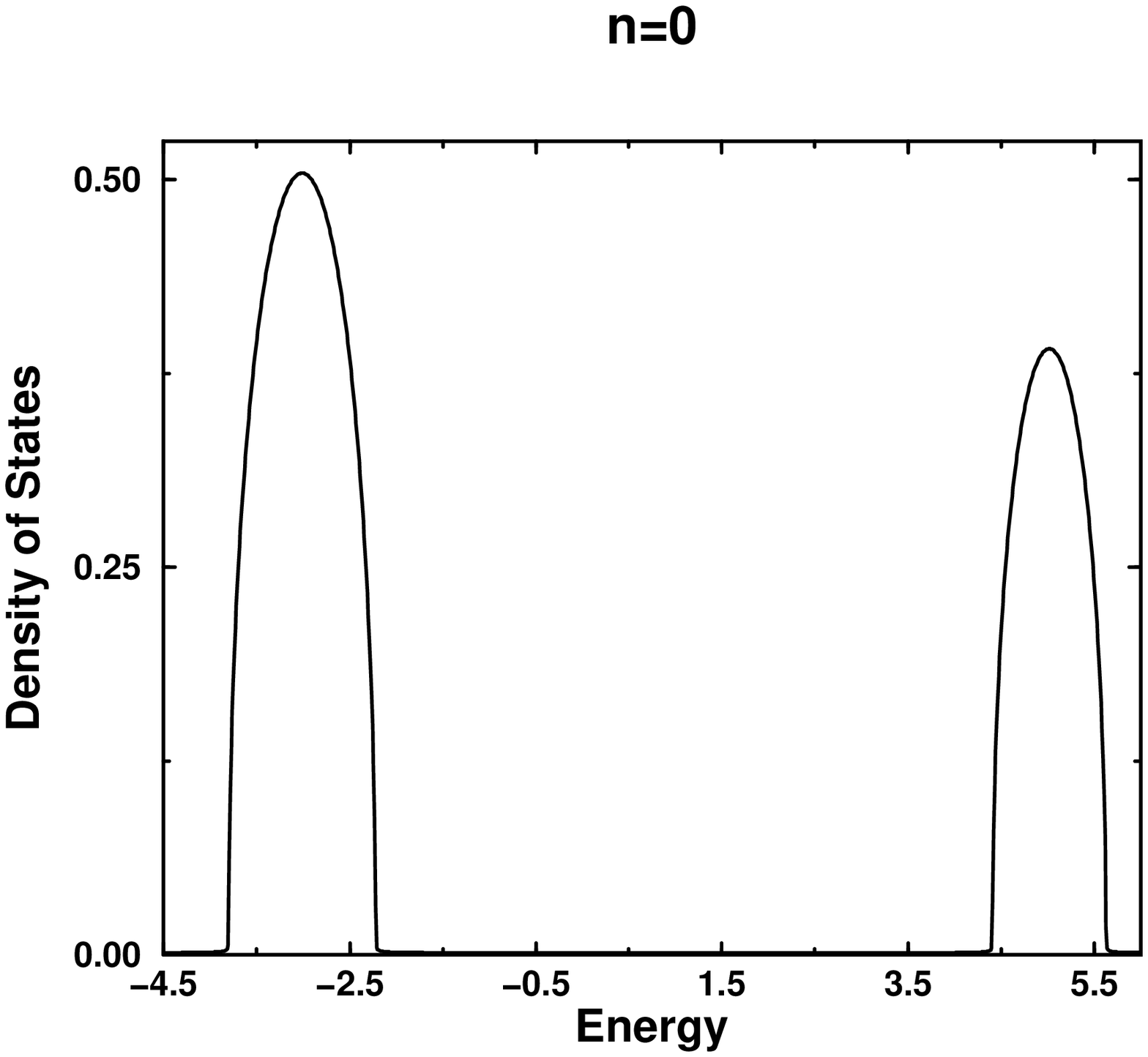}
  \end{minipage}}%
\subfigure[]{\label{fig:DOS_DE_model_b}
  \begin{minipage}[b]{0.5\textwidth}
    \centering \includegraphics[width=\textwidth]{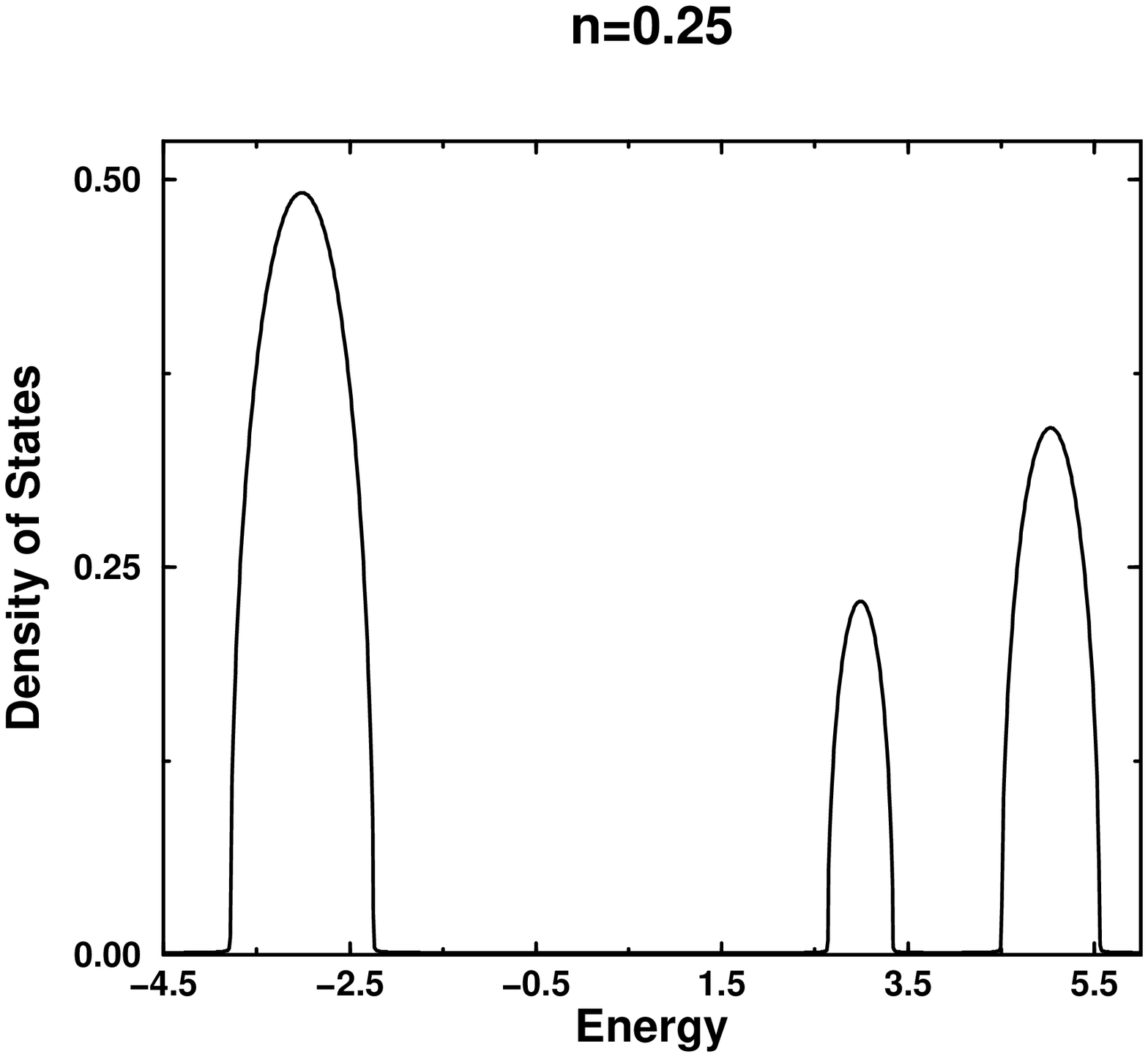}
  \end{minipage}}
\subfigure[]{\label{fig:DOS_DE_model_c}
  \begin{minipage}[b]{0.5\textwidth}
    \centering \includegraphics[width=\textwidth]{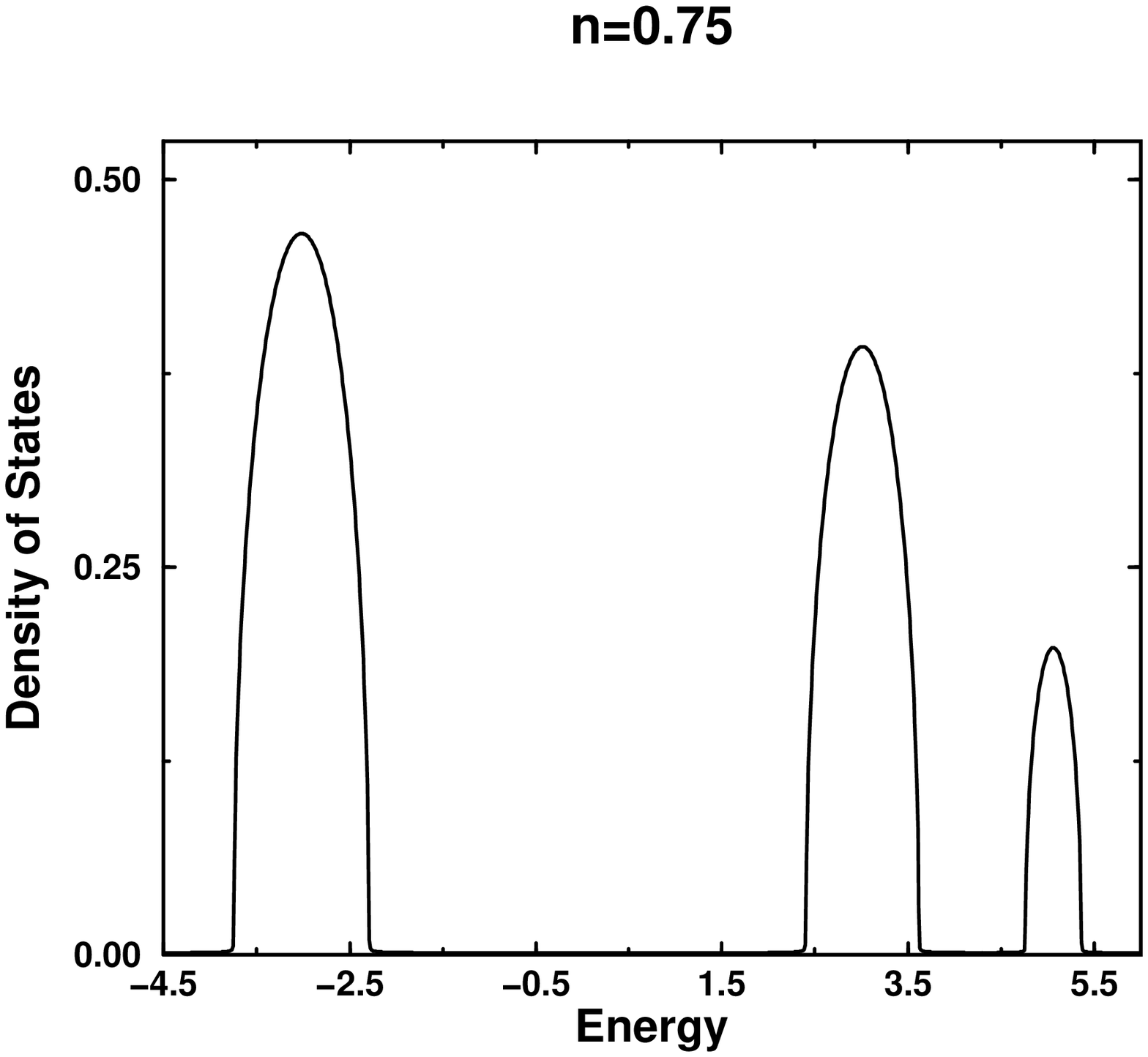}
  \end{minipage}}%
\subfigure[]{\label{fig:DOS_DE_model_d}
  \begin{minipage}[b]{0.5\textwidth}
    \centering \includegraphics[width=\textwidth]{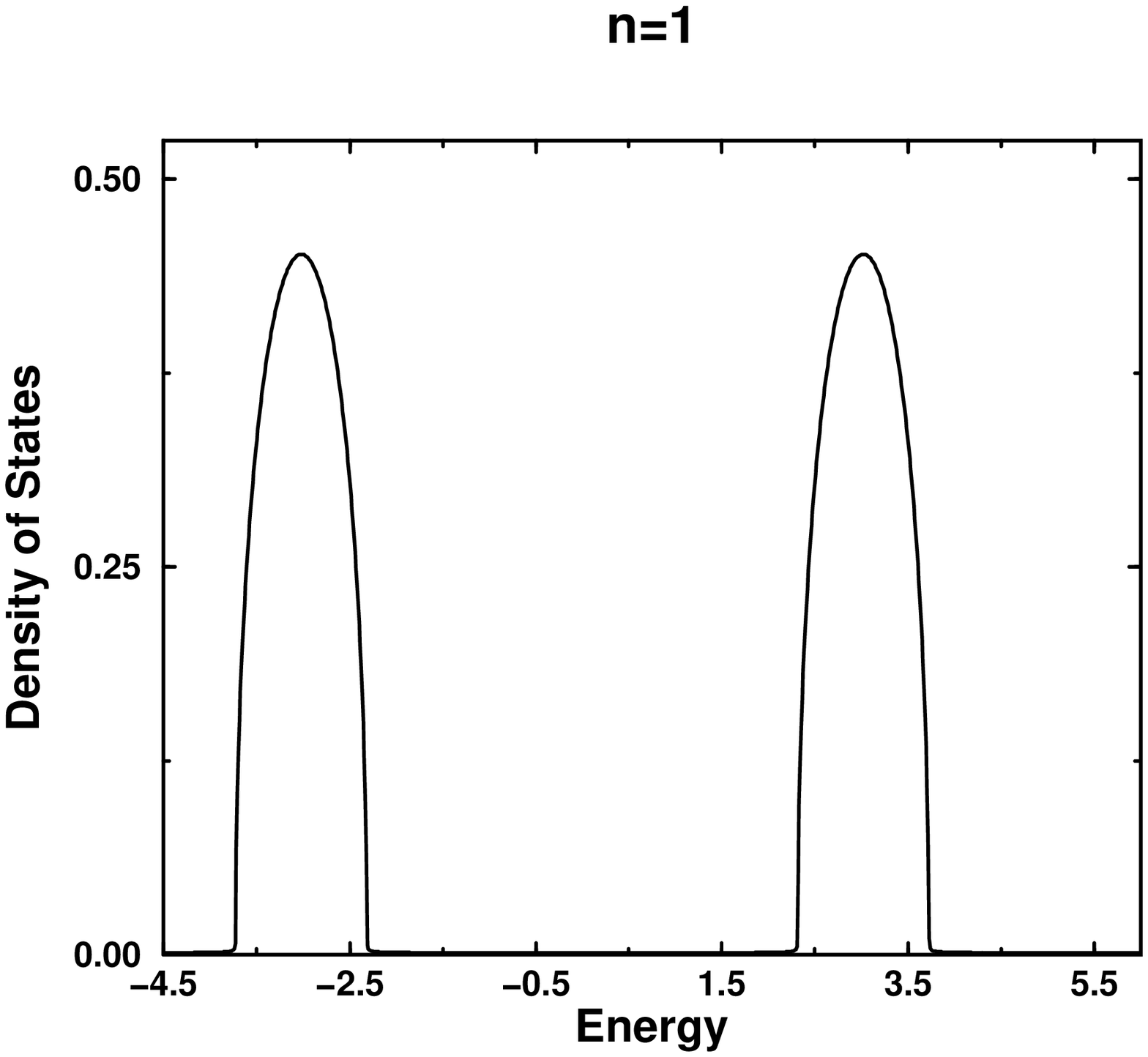}
  \end{minipage}}
  \caption{\label{fig:DOS_DE_model}%
    The density of states in the paramagnetic state of the double-exchange
    model for $S=3/2$, $J=4W$, and $n=0,\,0.25,\,0.75$ and $1$. Energy units
    of $W$ are used.}
\end{figure}

In the zero-field paramagnetic case it turns out that the CPA equation for
$G(\en)$ with the redundant suffix $\si$ omitted, is given by
equation~(\ref{eq:cpa5}) as in the Hubbard model. Thus, taking the band to
have the elliptic form (\ref{eq:ell_dos}), the CPA equation for $G$ is
\begin{equation}\label{eq:de_CPA}
  G(\en)=G^{\rm AL}\left(\en-W^2 G/4\right)
\end{equation}
with $G^{\rm AL}$ given by
\begin{equation}\label{eq:de_AL}
  \begin{split}
    G_\up^{\rm AL}(\en)=\frac{1}{2S+1}\left[\frac{nS/2-\la\bi{S}\cdot\bsigma\ra}
    {\en+J(S+1)/2}+\frac{S(1-n/2)-\la\bi{S}\cdot\bsigma\ra}{\en-J(S+1)/2}\right.\\
    \left.{}+\frac{(S+1)(1-n/2)+\la\bi{S}\cdot\bsigma\ra}{\en+JS/2}
    +\frac{n(S+1)/2+\la\bi{S}\cdot\bsigma\ra}{\en-JS/2}\right]\,.
  \end{split}
\end{equation}
The spin symmetry of the paramagnetic state has been used to simplify the
expectations in the previous form of $G^{\rm AL}$,
equation~(\ref{eq:de_AL_general}).  It is easy to show that
$\las\bi{S}\cdot\bsigma\ras\rightarrow nS/2$ as $J\rightarrow\infty$ and
$\las\bi{S}\cdot\bsigma\ras$ will be very near this limit as long as $JS\gtrsim
2W$. We make this approximation in calculating $G(\en)$, and hence the
density of states, $N(\en)=-\pi^{-1}\Im G(\en)$ from
equations~(\ref{eq:de_CPA}) and (\ref{eq:de_AL}). The results are shown in
figure~\ref{fig:DOS_DE_model} for $S=3/2$ and $J=4W$ for various $n$.
Clearly, from equation~(\ref{eq:de_AL}), the approximation to
$\las\bi{S}\cdot\bsigma\ras$ has the effect of removing the weak band centred
on $\en=-J(S+1)/2$ but it does not affect the total weight or the
distribution of weight between the two lower and two upper bands. It may be
seen that as $n$ increases from $0$ the band near $\en=J(S+1)/2$ is reduced
in weight and a new band appears near $JS/2$, until at $n=1$ no weight
remains in the band near $J(S+1)/2$. The weight in the band near $-JS/2$ is
$(S+1-n/2)/(2S+1)$ per spin so if $JS$ is sufficiently large to separate the
bands ($JS\gtrsim 2W$) this band will just be filled at $n=1$ producing a Mott
insulator as expected. This redistribution of weight between bands as they
fill with electrons is characteristic of the many-body CPA \cite{Ku74} and
was missing from Kubo's one-electron CPA which was restricted to $n=0$.

In the strong-coupling limit $J\rightarrow\infty$, which is taken with a
shift of energy origin $\en\rightarrow\en-JS/2$, equation~(\ref{eq:de_AL})
simplifies to
\begin{equation}\label{eq:de_AL_Jinf}
  G^{\rm AL}(\en)=\en^{-1}(S+1-n/2)/(2S+1)\,.
\end{equation}
Equation~(\ref{eq:de_CPA}) then becomes a quadratic equation for $G$ with
solution
\begin{equation}\label{eq:de_gf}
  G(\en)=\alpha^2\frac{2}{D^2}\left[\en-\sqrt{\en^2-D^2}\right]
\end{equation}
where $\alpha^2=(S+1-n/2)/(2S+1)$ and $D=\alpha W$. By comparing with
equations~(\ref{eq:ell_dos}) and (\ref{eq:G_0}) we see that the density of
states is a single elliptical band of weight $\alpha^2$ and bandwidth
$2\alpha W$. As $S\rightarrow\infty$ the band-narrowing factor
$\alpha\rightarrow 1/\sqrt{2}=0.707$, which is close to the classical result
of $2/3$, obtained by averaging $\cos\left(\theta/2\right)$ over the solid
angle.

In the classical spin limit $S\rightarrow\infty$ we rescale $J$, replacing it
by $J/S$, and the CPA equation becomes
\begin{equation}\label{eq:de_Sinf}
  G=\frac{1}{2}\left[\frac{1}{\Si+G^{-1}+J/2}+\frac{1}{\Si+G^{-1}-J/2}\right]\,.
\end{equation}
Here we have used the general equation~(\ref{eq:cpa5}), valid for arbitrary
band-shape, rather than equation~(\ref{eq:de_CPA}).
Equation~(\ref{eq:de_Sinf}) is precisely the equation obtained by Furukawa
\cite{phys_of_mns} within DMFT.


\section{Resistivity in the paramagnetic state of the DE model}\label{sec:resist-param-state}

The Kubo formula for the conductivity $\si$ involves the two-particle
current-current response function. However in the local approximation of CPA
or DMFT there is no vertex connection and $\si$ may be expressed in terms of
the one-particle spectral function
\begin{equation}\label{eq:spec_function}
  A_\bi{k}(\en)=-{\pi}^{-1}\Im G_\bi{k}(\en)=-{\pi}^{-1}\Im
  \left[\en-\en_\bi{k}-\Si(\en)\right]^{-1}\,.
\end{equation}
In the paramagnetic state $G$ is $T$-independent if we assume
$\las\bi{S}\cdot\bsigma\ras=nS/2$, and $\si$ depends on temperature only
weakly through the Fermi function. If we neglect this thermal smearing around
the Fermi energy $\mu$ we may calculate at $T=0$ but consider the results to
apply to the actual paramagnetic state at $T>\Tc$. We find \cite{EdGrKu99}
\begin{equation}\label{eq:si_1}
  \si=\frac{2\pi \rme^2}{3Na^3\hbar}\sum_\bi{k} \bi{v}_{k}^2
  \left[A_\bi{k}(\mu)\right]^2
\end{equation}
where $\bi{v}_\bi{k}=\nabla\en_\bi{k}$ is the electron velocity and $a^3$ is
the volume of the unit cell. Since $A_\bi{k}$ depends on $\bi{k}$ only
through $\en_\bi{k}$ we may define a function $\phi(\en)$ such that
$\phi'(\en_\bi{k})=\left[A_\bi{k}(\mu)\right]^2$. Hence the sum in
equation~(\ref{eq:si_1}) may be written as
\begin{equation}\label{eq:sum_1}
  \sum_\bi{k}\nabla\en_\bi{k}\cdot\nabla\phi(\en_\bi{k})=-\sum_\bi{k}
  \phi(\en_\bi{k})\nabla^2\en_\bi{k}\,,
\end{equation}
the last step following by means of Gauss's theorem. For a simple cubic
tight-binding band $\en_\bi{k}=-2t\sum_\beta\cos k_\beta a$, with $\beta$
summed over $x,y,z$, $\nabla^2\en_\bi{k}=-a^2\en_\bi{k}$. Then the summand in
equation~(\ref{eq:sum_1}) is a function of $\en_\bi{k}$ only and
equation~(\ref{eq:si_1}) becomes
\begin{equation}\label{eq:si_2}
  \si=\frac{2\pi \rme^2}{3 a\hbar}\int\,\rmd E EN_{\rm c}(E)\phi(E)
\end{equation}
where $N_{\rm c}(E)$ is the density of states for the simple cubic band. If
$N_{\rm c}(E)$ is replaced by a suitably-scaled Gaussian $N_{\rm
  g}(E)=(3/\pi)^{1/2}W^{-1}\exp\left[-3(E/W)^2\right]$, corresponding to an
infinite dimensional approximation, equation~(\ref{eq:si_2}) may be
simplified by integrating by parts:
\begin{equation}\label{eq:si_3}
  \si=\frac{\pi \rme^2 W^2}{9a\hbar}\int\,\rmd EN_{\rm g}(E)
  \left[A_E(\mu)\right]^2\,.
\end{equation}
Here $A_E(\mu)$ is defined by the right-hand expression in
equation~(\ref{eq:spec_function}) with $\en=\mu$, $\en_\bi{k}=E$.

\begin{figure}[htbp]
  \centering \includegraphics[width=0.75\textwidth]{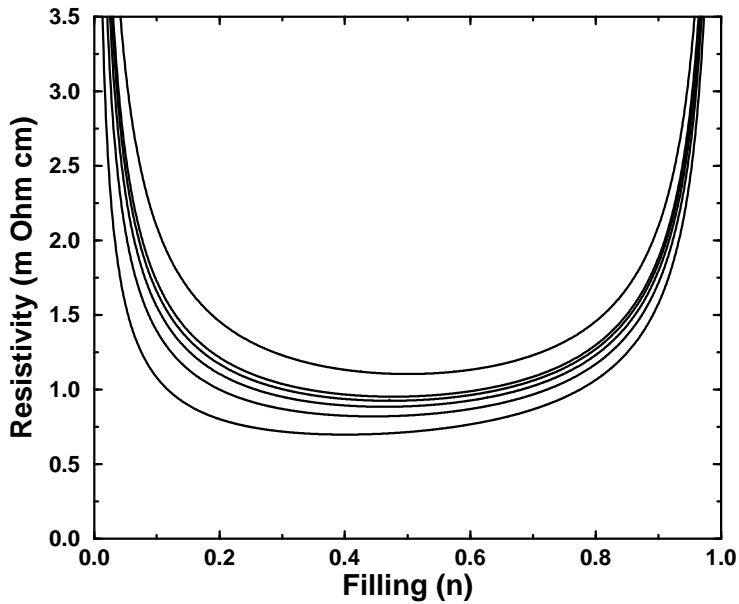}
    \caption{\label{fig:resis_elliptic}%
      The zero field paramagnetic state resistivity $\rho=\si^{-1}$ versus
      band-filling $n$ for the double-exchange model. Here $J=\infty$,
      $a=5{\rm\AA}$, and $S=1/2,\,1,\,3/2,\,2,\,5/2$ and $\infty$, $\rho$
      increasing with $S$. The elliptical density of states and
      formula (\ref{eq:si_2}) are used.}
\end{figure}

\begin{figure}[htbp]
  \centering \includegraphics[width=0.75\textwidth]{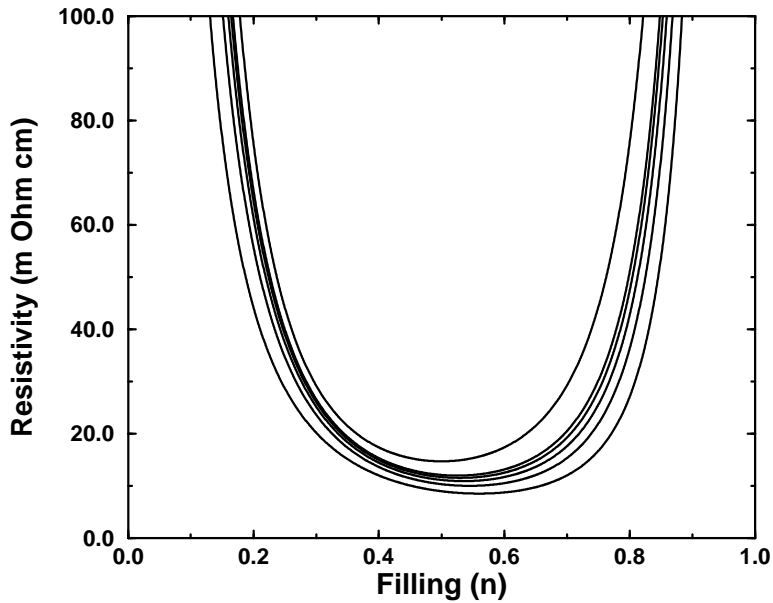}
    \caption{\label{fig:resis_lorentz}%
      As in figure~\ref{fig:resis_elliptic} but using a Lorentzian density of
      states and formula (\ref{eq:si_3}).}
\end{figure}

Since it is
convenient to use the elliptic density of states $N_0(\en)$ to calculate the
Green function and self-energy, as in the previous section, it is reasonable
to evaluate $\si$ using equation~(\ref{eq:si_2}) with $N_{\rm c}$ replaced by
$N_0$. We take the strong-coupling limit $J\rightarrow\infty$ for simplicity
and the results for the resistivity $\rho=\si^{-1}$ are plotted against
band-filling $n$ for various $S$ in figure~2. We have taken $a=5{\rm\AA}$
which is comparable with the \chem{Mn-Mn} spacing in manganites. For
$J=\infty$ the band-width $W$ is the only energy-scale and, since the
integral in equation~(\ref{eq:si_2}) is dimensionless, $\rho$ does not depend
on $W$. In fact in the DE regime $JS\gtrsim 2W$ the resistivity is almost
independent of both $J$ and $W$. Use of the alternative
formula (\ref{eq:si_3}), with $N_{\rm g}$ replaced by $N_0$, gives very
similar results. It is seen in figure~\ref{fig:resis_elliptic} that $\rho$
diverges correctly at $n=0$, owing to the absence of carriers, and at $n=1$
where the system becomes a Mott insulator. Well away from these insulating
limits $\rho$ does not depend strongly on $S$, so that quantum spin effects
are not very important. Furthermore $\rho\approx1{\rm m\Omega\,cm}$ over a
wide range of band-filling, which is much smaller than observed in some
manganites above $\Tc$, as discussed in section \ref{sec:appl-mang}. This
agrees with the conclusion of Millis \etal \cite{MiLiSh95} that the DE model,
with electrons scattered purely by disordered local spins, cannot describe
the physics of the manganites completely. Early work by Furukawa \cite{Fu94}
seemed to point to another conclusion, although the DMFT is equivalent to our
CPA approach. We showed \cite{EdGrKu99} that the confusion arose from
Furukawa's use of a convenient, but rather unreasonable, Lorentzian density
of states. Using a Lorentzian to replace $N_{\rm g}(E)$ and to calculate
$\phi(E)$ in equation~(\ref{eq:si_2}) leads to a divergent integral. However
equation~(\ref{eq:si_3}) is similar to the form of $\si$ used by Furukawa and
use of a Lorentzian to replace $N_{\rm c}(E)$ and calculate $A_E(\mu)$ gives
a convergent result.  Results of such calculations for the limit
$J\rightarrow\infty$ are shown in figure~\ref{fig:resis_lorentz} and it is
remarkable that $\rho$ is at least an order of magnitude larger than one
finds in figure~\ref{fig:resis_elliptic} for the more reasonable elliptic
band.


\section{Magnetism in the DE model}\label{sec:magnetism-de-model}

As discussed at the beginning of section \ref{sec:many-body-cpa}, the full
equation of motion approach to many-body CPA is required to determine
magnetic properties such as spin susceptibility $\chi$ and Curie temperature
$\Tc$. In our first paper \cite{EdGrKu99} this involved a hierarchy of Green
functions satisfying $4S+1$ coupled algebraic equations for local spin $S$;
only the $S=1/2$ case was briefly discussed. In a second paper \cite{GrEd99}
a major simplification was achieved by introducing generating Green functions
which generate all the required Green functions by differentiation with
respect to a parameter. The coupled equations are then replaced by a single
first-order linear differential equation, the parameter being the independent
variable, whose analytic solution yields the required CPA equations for the
Green functions. The classical limit $S=\infty$ can then be taken and for
$J=\infty$ the equations coincide with those of DMFT, which are only
obtainable in the classical limit. Our many-body CPA is therefore an analytic
approximation to DMFT for arbitrary quantum spin $S$ which becomes exact for
$S=\infty$. The many-body CPA also coincides with Kubo's \cite{Ku74}
one-electron CPA in the limit $n\rightarrow 0$ where that is valid.

\begin{figure}[htbp]
  \subfigure[]{\label{fig:Tc_n_elliptic}
  \begin{minipage}[b]{0.5\textwidth}
    \centering \includegraphics[width=\textwidth]{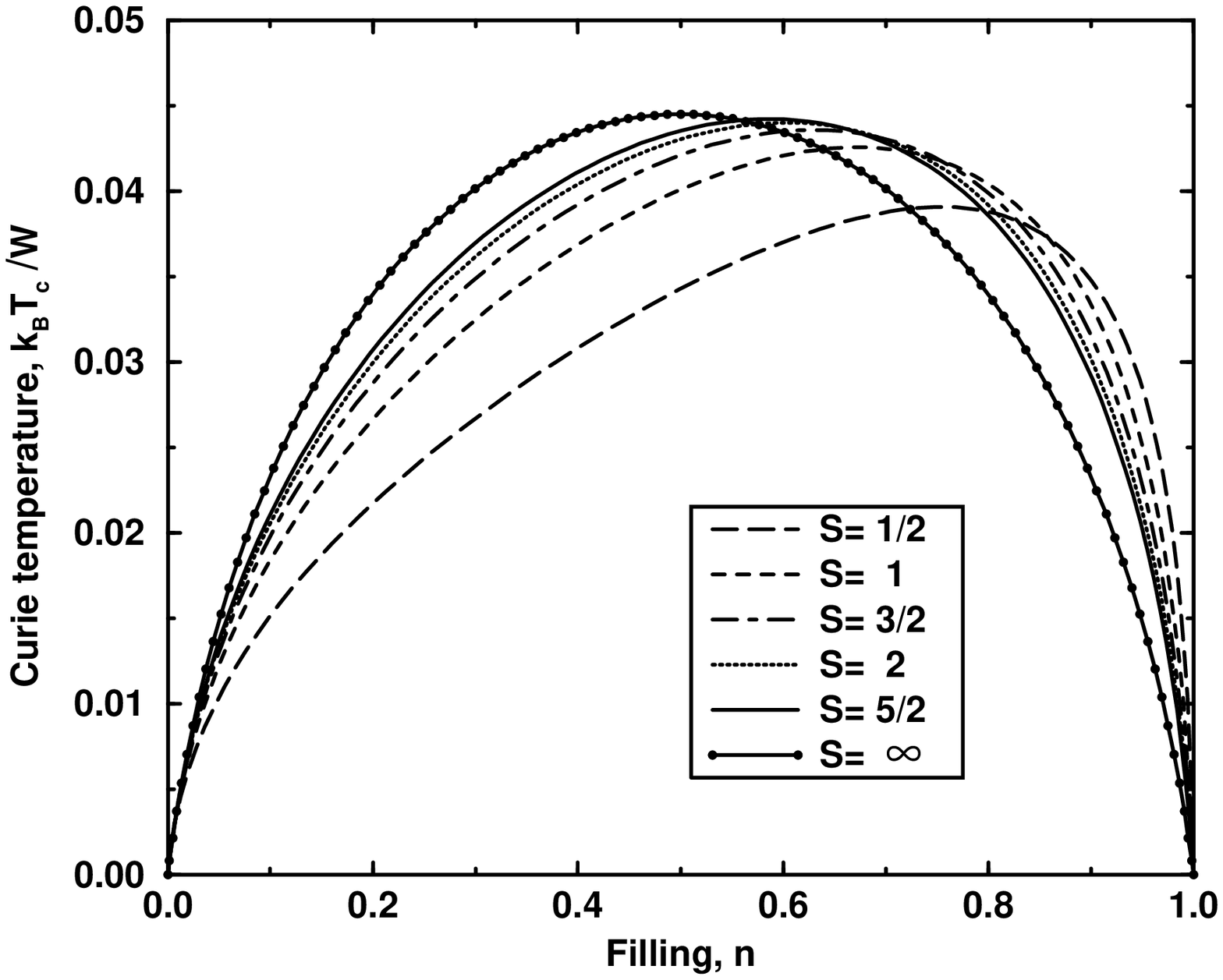}
  \end{minipage}}%
\subfigure[]{\label{fig:Tc_n_ell_cubic}
  \begin{minipage}[b]{0.5\textwidth}
    \centering \includegraphics[width=\textwidth]{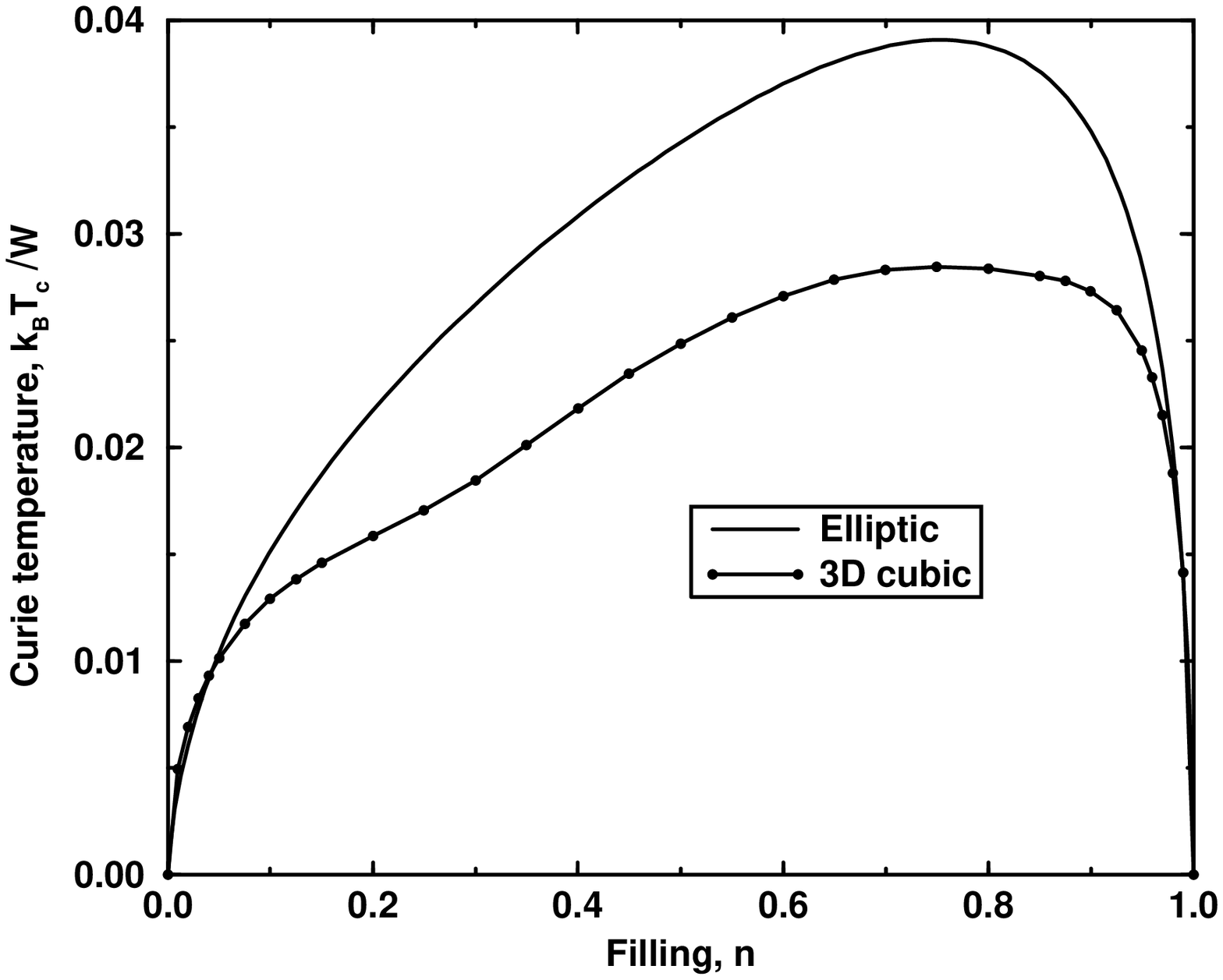}
  \end{minipage}}
  \caption{\label{fig:Tc_n}%
    The Curie temperature $k_{\rm B}\Tc/W$ of the double-exchange model
    versus band-filling $n$ for various $S$, calculated with $J=\infty$ using
    the elliptic band (a); the effect on $\Tc$ for $S=1/2$ of changing the
    elliptic band to the density of states for a simple cubic tight-binding
    band with nearest neighbour hopping (b).}
\end{figure}

To determine the magnetic properties one problem remains; the CPA and DMFT
equations contain one set of correlation functions $\las(S^z)^m\ras$ which
cannot be obtained directly from the Green functions. There is an indirect
procedure for determining these correlation functions within CPA but it
proves to be unsatisfactory, never yielding ferromagnetic solutions.
However, for $S=\infty$, DMFT provides a way to calculate the probability
distribution function $P(S^z)$, and hence $\las(S^z)^m\ras$, and we used an
empirical extension of this for finite $S$. This extension guarantees that
the spin susceptibility exhibits the correct Curie laws for band occupations
$n=0$ and $n=1$. Thus for $n=0$ we have a Curie law over the whole
temperature range, corresponding to $N$ independent spins $S$. For $n=1$,
with $J=\infty$, we have independent spins $S+\frac{1}{2}$. For $0<n<1$ we
find a finite Curie temperature $\Tc$ and some results \cite{GrEd99} are
shown in figure~\ref{fig:Tc_n}. In figure~\ref{fig:Tc_n_elliptic} $\Tc$ is
plotted as a function of $n$ for various $S$ with $J=\infty$, using the
elliptic band.  Clearly for finite $S$ ferromagnetism is more stable for
$n>0.5$ than for $n<0.5$, in agreement with the findings of Brunton and
Edwards \cite{BrEd98}. For $S=\infty$ the result agrees closely with that of
Furukawa \cite{phys_of_mns}. In figure~\ref{fig:Tc_n_ell_cubic} we see the
effect on $\Tc$ for $S=1/2$ of changing the bare band-shape from elliptic to
simple cubic tight-binding. A dip in $\Tc$ occurs around $n=0.3$ which is the
region where the ground state of the simple cubic DE model with $S=1/2$ is
rigorously not one of complete spin alignment \cite{BrEd98}.

\section{Many-body CPA for the Holstein-DE model}\label{sec:many-body-cpa-hde}

This section is based on Green's \cite{Gr01} recent study of the Holstein-DE
model in which the electrons of the DE model couple to local phonons as in
the Holstein treatment of small polarons \cite{Ho59a,Ho59b}. The Hamiltonian
is
\begin{equation}\label{eq:h_hde}
  \begin{split}
    H = &\sum_{ij\si} t_{ij}c_{i\si}^{\dagger}c_{j\si}-J\sum_i
    \bi{S}_i\cdot\bsigma_i-h\sum_i\left(S_i^z+\si_i^z\right)\\
    & -g\sum_i n_i\left(b_i^{\dagger}+b_i\right)+\om\sum_i
    b_i^{\dagger}b_i\,.
  \end{split}
\end{equation}
The first three terms constitute the DE Hamiltonian of
equation~(\ref{eq:h_de}) while the first, fourth and fifth terms form the
Holstein model. Einstein phonons on site $i$, with energy $\om$ and creation
operator $b_i^\dag$, couple to the electron occupation number $n_i=\sum_\si
n_{i\si}$ with coupling strength $g$. The electron-phonon coupling is of the
form $-g'\sum_i n_i x_i$, where $x_i$ is the displacement of a shell of atoms
surrounding site $i$, and in application to the manganites it may be regarded
as an effective Jahn-Teller coupling. Previous studies of this model have
either concentrated on coherent polaron bands, like R\"oder \etal
\cite{RoZaBi96}, or have treated the phonons classically \cite{MiMuSh96II} so
that there are no polaron bands at all. The many-body CPA approach is able to
encompass both aspects and to describe the crossover from quantum polarons to
the classical picture as temperature and/or model parameters are varied. The
relationship to previous theoretical work and to experimental studies of the
manganites is discussed fully in section \ref{sec:appl-mang}. However we
mention briefly below some related work on the pure Holstein model, without
coupling to local spins.

Sumi considered the Holstein model with one electron in the band, first
treating the phonons classically \cite{Su72} and later quantum mechanically
\cite{Su74}.  The classical case, with frozen displacements $x_i$,
corresponds to a multicomponent alloy for which CPA is the best local
approximation. In his dynamical CPA treatment of quantum phonons, Sumi
\cite{Su74} treated the one-site dynamics correctly and his work is
completely equivalent to the more recent DMFT treatment of Ciuchi \etal
\cite{CiPaFrFe97}. As a general rule dynamical CPA and DMFT are the same for
one-electron problems. DMFT is the correct extension of CPA to the many-body
problem of finite electron density but for the Holstein model, as for the DE
model, it cannot be carried through analytically in the quantum case.
Numerical work \cite{FrJaSc93,MiMuShI} applying DMFT to the Holstein model
has been aimed mostly at understanding superconducting transition
temperatures and charge-density-wave instabilities rather than the
polaron physics we are mainly concerned with. An unfortunate feature of the
Holstein model for spin $1/2$ electrons is that in a quantum treatment the
true ground state for strong electron-phonon coupling consists of unphysical
singlet bipolarons with two electrons bound on the same site. This problem
does not occur in the one-band Holstein-DE model since strong coupling $J$ to
local spins prevents double occupation of sites, as pointed out earlier. It
is also bypassed if the phonons are treated classically, as in the work of
Millis \etal \cite{MiMuShI} on the Holstein model. The Holstein model is more
complicated than the DE model and it turns out that our many-body CPA no
longer reduces to the correct one-electron dynamical CPA/DMFT
\cite{Su74,CiPaFrFe97} as band-filling $n\rightarrow 0$.  Although correct in
the atomic limit $t_{ij}=0$, our theory is clearly cruder for the Holstein
and Holstein-DE model than for the pure DE model.

We start by deriving the Green function for the Holstein-DE model in the
atomic limit. The Hamiltonian $H_{\rm AL}$ in this limit is given by
equation~(\ref{eq:h_hde}) with the first term omitted and with site indices
and summation suppressed.  We remove the electron-phonon coupling by the
standard canonical transformation \cite{Ma90} $\tilde{H}=\rme^s H_{\rm
  AL}\rme^{-s}$ where $s=-(g/\om)n(b^\dag-b)$. Under this transformation
$b\rightarrow b+(g/\om)n$ and the Hamiltonian separates into a fermionic and
bosonic component:
\begin{equation}\label{eq:hde_1}
  \tilde{H}=H_{\rm f}+H_{\rm b}
\end{equation}
\begin{equation}\label{eq:hde_2}
  H_{\rm f}=-J\bi{S}\cdot\bsigma-h\left(S^z+\si^z\right)-\left(g^2/\om\right)n^2
  \,,\,H_{\rm b}=\om b^\dag b\,.
\end{equation}
The transformation corresponds to a displacement of the equilibrium position
of the phonon harmonic oscillator in the presence of an electron and the
downward energy shift $g^2/\om$ is a polaron binding energy which we write as
$\lambda\om$, where $\lambda=g^2/\om^2$. If two electrons occupy the site
($n=2$), which will not occur for large $J$, the energy shift becomes
$4g^2/\om^2$ corresponding to an on-site bipolaron. Writing out explicitly
the thermal average in the definition of the one-particle retarded Green
function we have
\begin{eqnarray}\label{eq:hde_3}
  G_\si^{\rm AL}(t)&=&-\rmi\theta(t)\la\left[c_\si(t),c_\si^\dag\right]_+
  \ra\nonumber\\
  &=&-\rmi\theta(t)\frac{\Tr\left\{\rme^{-\beta H_{\rm AL}}\left[c_\si(t),
  c_\si^\dag\right]_+\right\}}{\Tr\left\{\rme^{-\beta H_{\rm AL}}\right\}}
\end{eqnarray}
and the canonical transformation introduced above can be carried out within
the traces, using the property of cyclic invariance. Thus $H_{\rm
  AL}\rightarrow\tilde{H}$, $c_\si^\dag\rightarrow X^\dag c_\si^\dag$ and
$c_\si(t)$ becomes
\begin{equation}\label{eq:hde_4}
  \rme^{\rmi\tilde{H}t}Xc_\si \rme^{-\rmi\tilde{H}t}
\end{equation}
where $X=\exp\left[g(b^\dag-b)/\om\right]$. Using equation~(\ref{eq:hde_1}),
we can write the traces in equation~(\ref{eq:hde_3}) as products of fermionic and
bosonic traces. Hence we find
\begin{equation}\label{eq:hde_5}
  G_\si^{\rm AL}(t)=-\rmi\theta(t)\left\{\la c_\si(t)c_\si^\dag\ra_{\rm f}F(t)+
    \la c_\si^\dag c_\si(t)\ra_{\rm f} F^*(t)\right\}
\end {equation}
where $F(t)=\las X(t)X^\dag\ras_{\rm b}$ and the thermal averages
$\las\ \ras_{\rm f}$, $\las\ \ras_{\rm b}$ correspond to the systems with
Hamiltonians $H_{\rm f}$ and $H_{\rm b}$ respectively. It may be shown \cite{Ma90}
that
\begin{equation}\label{eq:hde_6}
  F(t)=\rme^{-\lambda(2b+1)}\exp\left\{2\lambda\left[b(b+1)\right]^{1/2}\cos
    \left[\om(t+i\beta/2)\right]\right\}
\end{equation}
where $b=b(\om)=\left(\rme^{\beta\om}-1\right)^{-1}$ is the
Bose function with $\beta=\left(k_{\rm B}T\right)^{-1}$. The last factor is
of the form $\exp\left(z\cos\phi\right)$
which generates the
modified Bessel functions ${\rm I}_r(z)$:
\begin{equation}\label{eq:hde_7}
  \exp\left(z\cos\phi\right)=\sum_{r=-\infty}^{\infty}{\rm I}_r(z)\rme^{\rmi r\phi}\,.
\end{equation}
To evaluate the fermionic averages we
consider for simplicity the limit $J\rightarrow\infty$ in zero field
($h=0$). Then the last term in $H_{\rm f}$ may
be written $-(g^2/\om)n$, since $n=0$ or $1$ only, and this may be
absorbed into the chemical
potential which is finally determined to give the correct number of electrons
$n$ per atom. Thus $H_{\rm f}$ is just the DE Hamiltonian in the atomic limit
and the sum of
the two fermionic averages corresponds to the function $G^{\rm AL}(t)$ whose
Fourier
transform is given by equation~(\ref{eq:de_AL_Jinf}). It is easy to see that the
first and second
thermal averages in equation~(\ref{eq:hde_5}) take constant values
$(1-n)(S+1)/(2S+1)$ and
$n(S+1/2)(2S+1)$ respectively. Hence, from
equations~(\ref{eq:hde_5})-(\ref{eq:hde_7}), we obtain the
Fourier transform of $G^{\rm AL}$, with $J\rightarrow\infty$ and $h=0$, in
the form 
\begin{equation}\label{eq:G_hde_AL}
  G^{\rm AL}(\en)=\sum_{r=-\infty}^{\infty}\frac{{\rm I}_r\{2\lambda
      \left[b(\om)(b(\om)+1)\right]^{1/2}\}}
  {(2S+1)\exp\{\lambda\left[2b(\om)+1\right]\}}
  \frac{(2S+1)\frac{n}{2}\rme^{r\beta\om/2}+(S+1)(1-n)\rme^{-r
      \beta\om/2}}{\en+r\om}\,.
\end{equation}
\begin{figure}[htbp]
  \centering \includegraphics[width=0.75\textwidth]{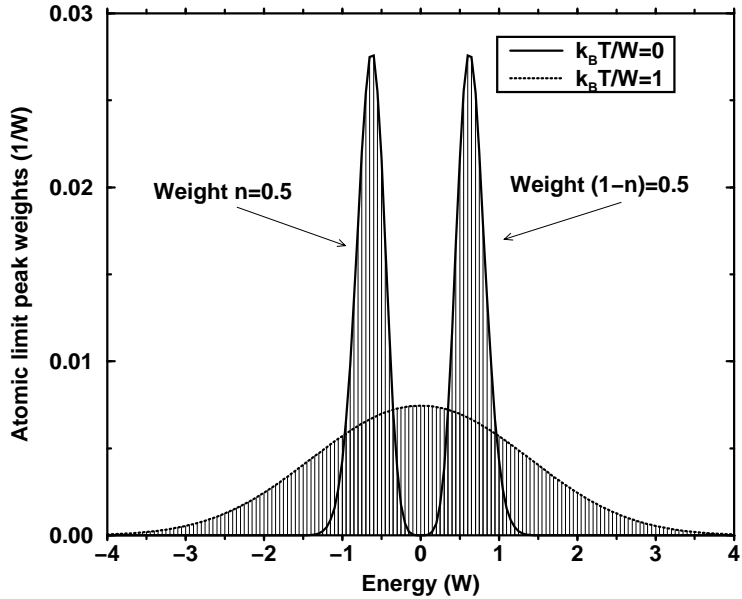}
    \caption{\label{fig:hde_AL}%
      One-electron spectra of the Holstein-DE model in the atomic limit at
      zero and very high temperature. They consist of delta-functions, with
      energy spacing $\om$, whose strength is indicated by the envelope
      curves. The plots are for the paramagnetic state with $S=J=\infty$,
      $h=0$, $n=0.5$, $\om/W=0.05$ and $g/W=0.18$, where $W$ is a unit of
      energy later to be identified with the half-width of the electron band
      in the full Hamiltonian.}
\end{figure}
The density of states $-\pi^{-1}\Im\,G^{\rm AL}(\en)$ is shown in
figure~\ref{fig:hde_AL} for
the classical spin limit $S\rightarrow\infty$ at
quarter-filling $n=0.5$. It consists of delta-function peaks separated in
energy by $\om$
and the envelope curves show the weight distribution at low and high
temperature. $W$ is an energy unit which, when we go beyond the atomic limit,
will be the half-width of the itinerant electron band, as usual. The values
adopted for the parameters $\om/W$ and $g/W$ relate to the manganites, as
discussed in section \ref{sec:appl-mang}. The symmetry of the spectrum about
zero energy is due to the
choice of filling $n=0.5$; in general at $T=0$ the lower and upper ``bands''
have weights $n$
and $1-n$ respectively. By counting weights it may be seen that for any $n$ the
chemical potential lies in the peak at $\en=0$, which has very small weight
$\rme^{-\lambda}/2$ per
spin. The shape of the envelope function at $T=0$, with two maxima and very small
values at the centre of the pseudogap between them, may be understood
physically as follows. The delta-function at $\en=r\om$ ($r\ge 0$)
corresponds to an excitation
from the ground state, with no electron and the undisplaced oscillator in its
ground state, to a state with one electron and the displaced oscillator in
its $r^{\rm th}$ excited state. The strength of the delta-function depends on
the square
of the overlap integral between the displaced and undisplaced oscillator wave
functions. Clearly this is very small for $r=0$ and goes through a maximum with
increasing $r$ as the normalized displaced wave function spreads out. At
$T=0$ it is
easily seen from equation~(\ref{eq:G_hde_AL}), using ${\rm I}_r(z)\sim
(z/2)^{|r|}/|r|!$ for small
$z$, that the weight of the
delta-functions at $\en=\pm r\om$ is proportional to $\lambda^r/r!$ . Hence
the maxima in the envelope curve occur at $\en\approx\pm\lambda\om$, which
is the polaron binding energy.

We now turn to the
Holstein-DE model with finite band-width. As for the DE model it is 
necessary to use the full equation~of motion method to derive the many-body 
CPA in the presence of a magnetic field and/or magnetic order \cite{Gr01}. In the 
present case it is very difficult to determine self-consistently all the 
expectation values which appear. We therefore approximate them by their 
values in the atomic limit. It then turns out that in the zero field 
paramagnetic state, for $J=\infty$ and with the elliptic band, the CPA Green
function $G$ again satisfies equation~(\ref{eq:de_CPA}), with $G^{\rm AL}$ now given
by equation~(\ref{eq:G_hde_AL}).

\begin{figure}[htbp]
  \centering \includegraphics[width=0.75\textwidth]{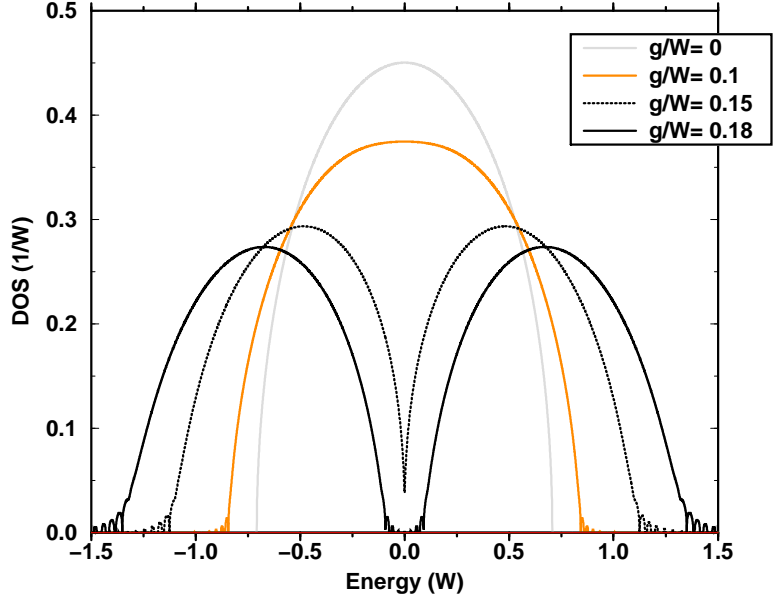}
    \caption{\label{fig:DOS_g}%
      The one-electron density of states (DOS) for the Holstein-DE model with
      half-bandwidth $W$, for the hypothetical paramagnetic state at $T=0$,
      with various strengths of electron-phonon coupling $g/W$.  Other
      parameters as in figure~\ref{fig:hde_AL}.}
\end{figure}

\begin{figure}
  \centering \includegraphics[width=0.75\textwidth]{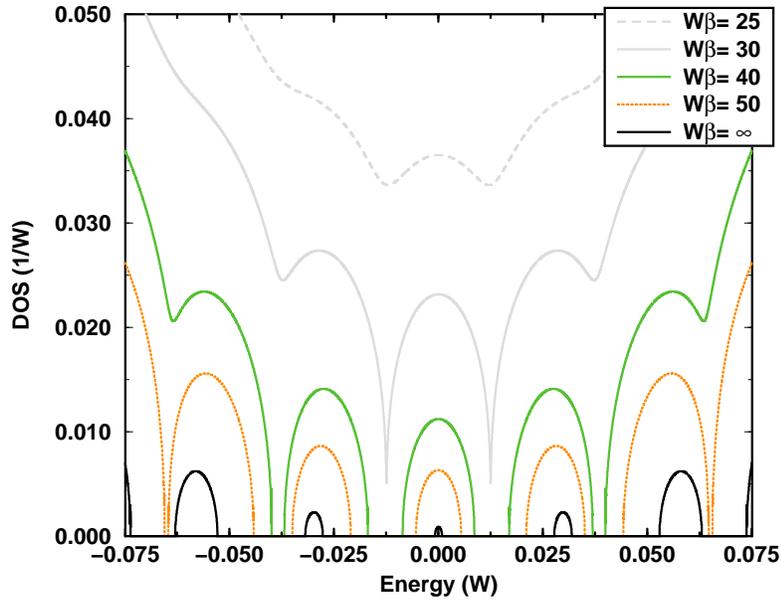}
    \caption{\label{fig:DOS_polarons}%
      Evolution with temperature $\beta=\left(k_{\rm B}T\right)^{-1}$ of the
      polaron subbands in the pseudogap around the chemical potential $\mu=0$
      for $g/W=0.18$. These subbands at $T=0$ can just be seen in
      figure~\ref{fig:DOS_g}. All parameters as in figure~\ref{fig:hde_AL}.}
\end{figure}

The densities of states calculated for $T=0$ using equations~(\ref{eq:G_hde_AL})
and (\ref{eq:de_CPA}) with $S=\infty$, $n=0.5$, $\om/W=0.05$ and various
values of $g/W$ are
shown in figure~\ref{fig:DOS_g}. Apart from lacking the perfect symmetry about
the chemical
potential $\mu=0$ the results are qualitatively similar for other values of
$n$ not too close to $0$ or $1$. For $g=0$ we recover the elliptic band with
half-width $W/\sqrt{2}$ as for the DE model with $J=\infty$, $S=\infty$. As
$g$ increases the density of states broadens and small
subbands are split off from the band edges. As $g$ increases further a pseudogap
develops near the chemical potential. At a critical value $g=g_{\rm c}$ a gap
appears
which contains a small polaron band around the chemical potential.
Increasing $g$ further causes more bands to be formed in the gap, with weights
similar to those of the relevant atomic limit. It should be pointed out that
the paramagnetic state considered here at $T=0$ is not the actual ground state,
which is ferromagnetic. We discuss the magnetic state later. The effect of
increasing temperature on the density of states in the gap region is shown in
figure~\ref{fig:DOS_polarons} for $g=0.18W>g_{\rm c}$. With increasing $T$ the
polaron bands grow
rapidly and eventually merge to fill the gap.

It is important to compare these results with the standard small
polaron theory developed by Holstein [25, 33]. Holstein distinguished
between ``diagonal transitions'', in which the number of phonons is unchanged
as the electron moves from site to site, and ``nondiagonal transitions'' in
which phonon occupation numbers change. The former give rise to a coherent
Bloch-like polaron band of half-width $W\rme^{-\lambda(2b+1)}$ which decreases
with increasing
temperature. The nondiagonal transitions are inelastic processes which
destroy phase coherence and the polaron moves by diffusive hopping. The
hopping probability increases with temperature so that polaron motion crosses
over from coherent Bloch-like at $T=0$ to diffusive hopping as $k_{\rm B}T$
approaches the
phonon energy $\om$. The paramagnetic state of the Holstein-DE model
differs from
this standard picture in one important respect. There are no well-defined
Bloch states, owing to strong scattering by the disordered local spins, so no
coherent polaron band will form. This is fortunate because the CPA treatment
of electron-phonon scattering will never lead to coherent states of infinite
lifetime at the Fermi surface at $T=0$. However in the presence of strong spin
disorder it should be satisfactory. We interpret the central band around the
chemical potential in figure~\ref{fig:DOS_polarons} as an incoherent polaron
band whose increasing
width as the temperature rises is due to life-time broadening of the atomic
level. The life-time decreases as the hopping probability increases with
rising temperature.

To substantiate this picture we study the central
polaron band in the limit of very strong electron-phonon coupling. In this
limit it can be shown that we need retain only the $r=0$ term in
equation~(\ref{eq:G_hde_AL}) and it is
then easy to solve equation~(\ref{eq:de_CPA}) for $G$. The result is of the same
form as equation~(\ref{eq:de_gf}) but with
\begin{equation}\label{eq:hde_8}
  D^2=\frac{1}{2}W^2\rme^{-\lambda(2b+1)}{\rm
  I}_0\left(2\lambda\left[b(b+1)\right]^{1/2}\right)\,.
\end{equation}
The central band is thus elliptical with half-width $D$ and
weight $D^2/W^2$. It is now easy to calculate the conductivity $\si$ from
equation~(\ref{eq:si_1}) and, using $D^2\ll W^2$, we find
\begin{equation}\label{eq:hde_9}
  \si=\frac{\pi \rme^2}{6\hbar a}\frac{D^2}{W^2}\approx\frac{\pi \rme^2}{12\hbar a}
  \left(\frac{\beta\om}{4\pi\lambda}\right)^{1/2}\rme^{-\beta\lambda\om/4}\,.
\end{equation}
The last step follows by using the asymptotic forms for
strong coupling and high temperature ${\rm I}_0(z)\sim\left(2\pi
  z\right)^{-1/2}\exp{z}$ and $b\sim\left(\beta\om\right)^{-1}$. The
temperature dependence of $\si$ is
the same as in the standard theory of small polaron hopping conduction \cite{Ma90}
but with activation energy $\lambda\om/4$ equal to one quarter, instead of
one half, of the polaron binding energy. The standard result is for one
electron coupled to
the lattice without spin disorder. Nevertheless this establishes the link
between our work and standard small polaron theory in the strong coupling
limit. However the results shown in figure~\ref{fig:DOS_polarons}, with
parameters relevant to
typical manganites, are far from this limit. They correspond to intermediate
coupling and in the actual paramagnetic state above the Curie temperature the
polaron bands are largely washed out. In this regime, with increasing
temperature, there is a crossover from polaronic behaviour to a situation
where the phonons behave classically, the case considered by Millis \etal
\cite{MiMuSh96II}. For electron-phonon coupling greater than a critical value these
authors find a gap in the density of states which gradually fills with
increasing temperature. However in their classical treatment there are no
polaron bands in the gap so that the link with standard polaron physics is
not established.

Apart from the symmetry about $\en=0$ the above results for $n=0.5$ are not
untypical of the general case. For general $n$ the main lower and upper bands,
separated by a gap for $g>g_{\rm c}$, have approximate weights $n$ and $1-n$
respectively. The chemical potential at $T=0$ is always confined to the
polaron band arising from the $r=0$ term of equation~(\ref{eq:G_hde_AL}), and
moves from the bottom at $n=0$ to the top at $n=1$, so that we correctly have
an insulator in these limits.

\begin{figure}
  \centering \includegraphics[width=0.75\textwidth]{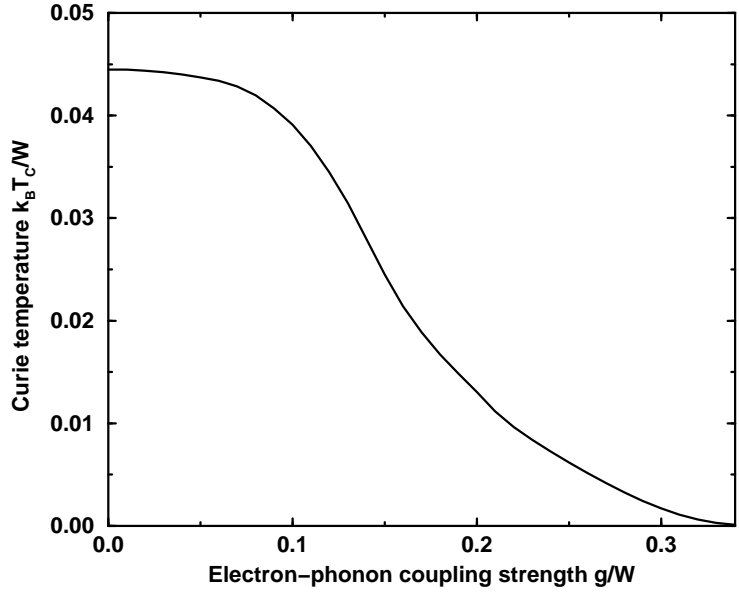}
    \caption{\label{fig:Tc_coupling}%
      Suppression of the Curie temperature of the Holstein-DE model with
      increasing electron-phonon coupling $g/W$. The plot is for
      $S=J=\infty$, $h=0$, $n=0.5$ and $\om/W=0.05$.}
\end{figure}

\begin{figure}
  \centering \includegraphics[width=0.75\textwidth]{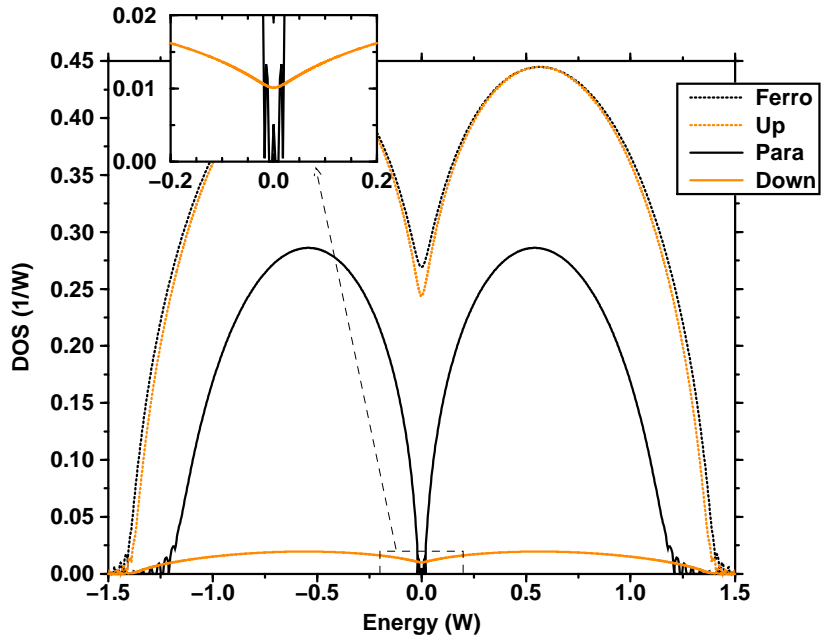}
    \caption{\label{fig:DOSs_PM_FM}%
      The up- and down-spin density of states of the Holstein-DE model with
      $g/W=0.16$ for $k_{\rm B}T=0.005W\ll k_{\rm B}\Tc$ where $\las
      S^z\ras=0.915$. Also shown are the DOS for the saturated ferromagnetic
      state at $T=0$ and for the hypothetical paramagnetic state at $T=0$.
      All plots are for $S=J=\infty$, $h=0$, $n=0.5$, $\om/W=0.05$ and
      $g/W=0.16$.}
\end{figure}

To calculate the Curie temperature $\Tc$ we need the full CPA theory
combined with an exact result of DMFT for $S=\infty$ \cite{Gr01}. Results on $\Tc$
for the same parameters as before are plotted as functions of electron-phonon
coupling $g$
in figure~\ref{fig:Tc_coupling}. The suppression of $\Tc$ with increasing $g$
was first noted by R\"oder \etal
\cite{RoZaBi96} and our own results are quite similar to those of Millis
\etal \cite{MiMuSh96II}. In our CPA we have no reliable means of calculating
the probability distribution
function $P(S^z)$, so to go below $\Tc$ we use the mean-field approximation for the
ferromagnetic Heisenberg model with classical spins and nearest neighbour
exchange. The exchange constant is determined by $\Tc$.
We plot the up- and
down-spin density of states for $T=0.005W/k_{\rm B}\ll\Tc$ and $g=0.16W>g_{\rm
  c}$ in figure~\ref{fig:DOSs_PM_FM}, also showing curves for the
saturated ferromagnetic state and paramagnetic state at $T=0$ for comparison. The
value $g=0.16W$ is closer to $g_{\rm c}$ than the value of $0.18W$ used in
figures~\ref{fig:DOS_g} and \ref{fig:DOS_polarons} and we
discuss these results in relation to the manganites in the next section. In
figure~\ref{fig:DOSs_PM_FM} we plot the resistivity $\rho$ as a function of
temperature, for the same parameter set, with different applied fields
$h$. The resistivity peaks sharply
at $\Tc$, and for comparison we show results for weaker electron-phonon
coupling $g/W=0.10$ in figure~\ref{fig:rho_T_g0.1}. The curve in
figure~\ref{fig:rho_T_g0.1} is almost indistinguishable from that of
figure~7 in reference~\cite{Gr01} for $g/W=0.01$. This
is not surprising since we see from figures~\ref{fig:DOS_g} and
\ref{fig:Tc_coupling} of this paper that the density of states and $\Tc$
change very little between $g/W=0$ and $g/W=0.1$. These results are all
discussed further in the next section.
\begin{figure}
  \centering \includegraphics[width=0.75\textwidth]{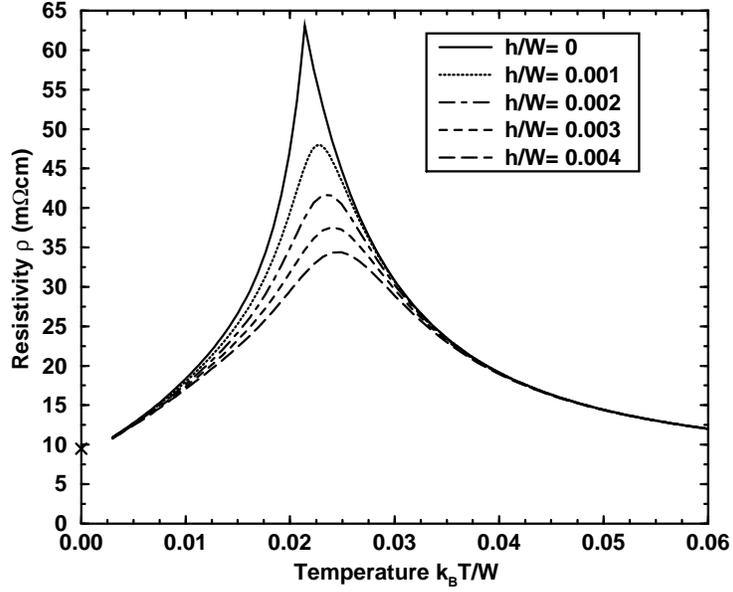}
    \caption{\label{fig:rho_T_g0.16}%
      Resistivity $\rho$ versus temperature for the Holstein-DE model with
      $S=J=\infty$, $n=0.5$, $\om/W=0.05$, intermediate coupling $g/W=0.16$
      and various applied fields $h$. The lattice constant is taken as
      $a=5{\rm\AA}$, slightly larger than the \chem{Mn-Mn} spacing in the
      manganites.}
\end{figure}
\begin{figure}
  \centering \includegraphics[width=0.75\textwidth]{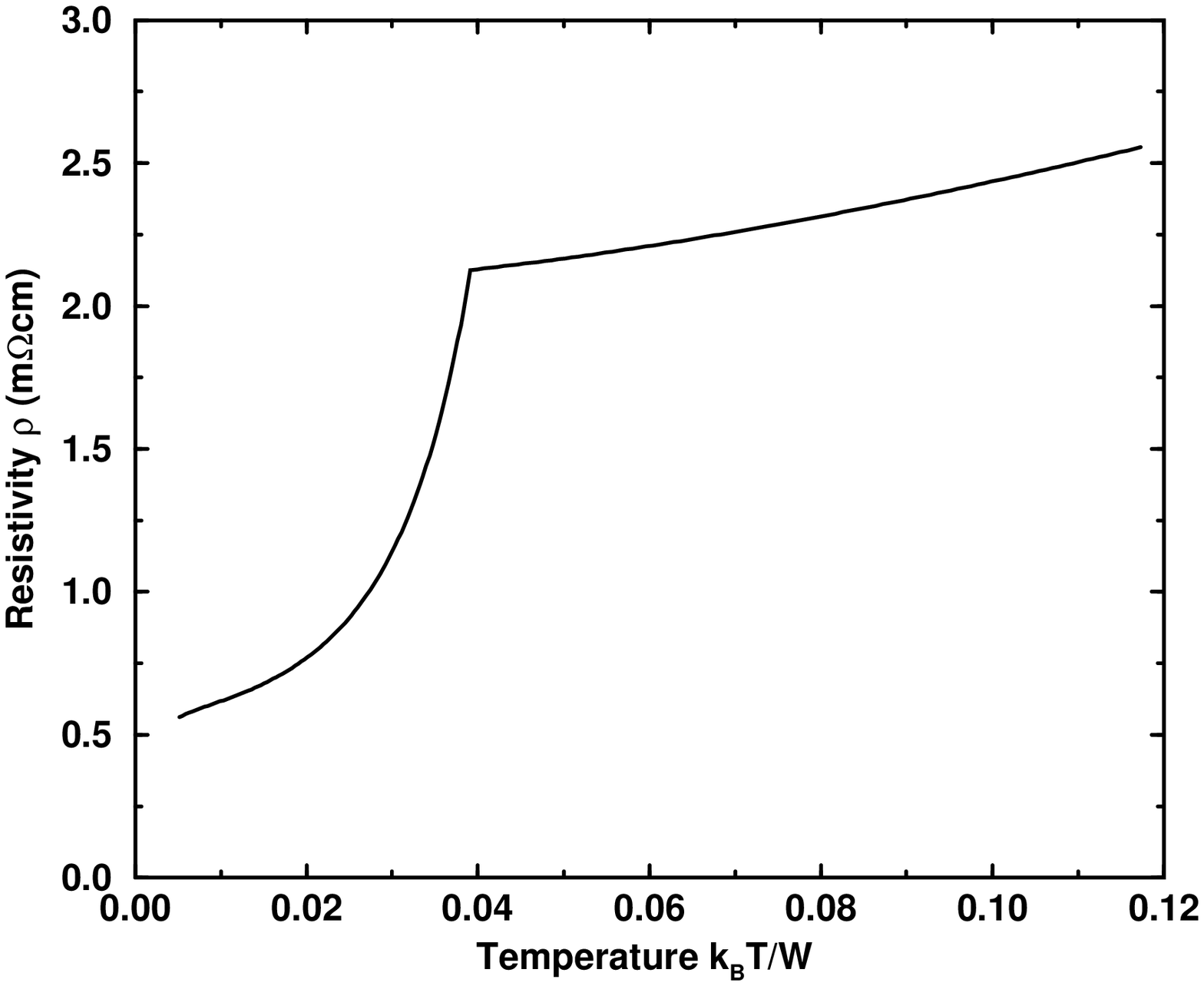}
    \caption{\label{fig:rho_T_g0.1}%
      The same plot as figure~\ref{fig:rho_T_g0.16} but for weak
      electron-phonon coupling $g/W=0.10$.}
\end{figure}


\section{Application to manganites}\label{sec:appl-mang}

Much experimental work has concentrated on the systems \chem{La_{1-x}Ca_x
  MnO_3} and \chem{La_{1-x}Sr_x MnO_3} with $x\approx0.33$, in the middle of
the ferromagnetic regime where CMR is observed. For brevity we denote these
systems by \chem{LCMO} and \chem{LSMO} respectively. In comparing experiments
with the detailed results of section \ref{sec:many-body-cpa-hde} it should be
borne in mind that the deduced parameters will be influenced by our
convenient choice of $n=1-x=0.5$ rather than $n=0.6-0.7$.  However the
correct general picture should emerge. Since we consider a homogeneous state
we are not concerned with the existence of charge ordering for $x=0.5$. We
should also discuss the effect of using a one-band model for the $e_{g}$
band, rather than the more realistic two-band model. An important point to
notice is that in the two-band model large Hund's rule coupling $J$ to the
local spins no longer produces a Mott insulator for $n=1$ and it is necessary
to introduce on-site Coulomb interaction \cite{BeZe99,HeVo00}. Millis \etal
\cite{MiMuSh96II} do not do this so that for $n=1$, which should correspond
to the undoped insulator, they have a metal. Also, for weak electron-phonon
interaction, they find $\Tc$ is largest for this value of $n$. This contrasts
strongly with figure~\ref{fig:Tc_n} where, in the one-band DE model, $\Tc=0$
at $n=0$ and $1$ and has a maximum in between. Held and Vollhardt
\cite{HeVo00}, using DMFT, obtain very similar results to ours for $S=\infty$
in the two-band model with strong on-site Coulomb interaction included.
However their values of $\Tc$ are about twice ours. Nevertheless we conclude
that the one-band DE and Holstein-DE models may sensibly be used to discuss
the manganites. We see from figures~\ref{fig:resis_elliptic}
and \ref{fig:Tc_n} that for the DE model neither $\Tc$ nor the resistivity
$\rho$ vary enormously with $S$ so that $S=\infty$ is a reasonable
approximation to the $S=3/2$ \chem{Mn} spin. It is also reasonable to take
the DE limit $J\rightarrow\infty$ \cite{Sa96}. For comparison with experiment
we take $W=1\,{\rm eV}$ which band calculations \cite{PiSi96,Sa96} suggest as
an appropriate half-bandwidth for the $e_{g}$ band. Also in figures~6-11
we have taken $\om/W=0.05$ to correspond to observed transverse optic phonons
with $\om\approx40-70{\rm meV}$ which couple strongly to the electrons in
\chem{LCMO} \cite{KiGuChPa96}.

Perhaps the most striking feature of the manganites is the very different
behaviour observed in apparently similar materials such as \chem{LSMO} and
\chem{LCMO}. For \chem{LSMO}, with $x\approx 0.33$, $\Tc\approx 370\,{\rm K}$
whereas for \chem{LCMO}, with a similar $x$, $\Tc\approx 240\,{\rm K}$. The
difference in behaviour of the resistivity $\rho$ above $\Tc$ is much more
striking. For \chem{LSMO} $\rho\approx 4\,{\rm m\Omega\,cm}$ and increases
slowly with temperature as in a poor metal \cite{UrMoAr95}. The $\rho(T)$
curve is very similar to that of figure~\ref{fig:rho_T_g0.1} for $g/W=0.1$
except for a much larger resistivity at low temperature in our calculations.
Since this feature persists even for $g/W=0.01$ (figure~7 in
reference~\cite{Gr01})
it presumably arises from overestimated spin disorder scattering
at low temperatures due to our use of the classical spin Heisenberg model to
determine $P(S^z)$. In \chem{LCMO} the resistivity rises to a maximum at
$\Tc$ of about $40\,{\rm m\Omega\,cm}$ and then falls with increasing
temperature above $\Tc$ \cite{McWaZh96}. In contrast to \chem{LSMO} there is
thus a transition from metallic to insulating behaviour. Also the resistivity
peak is strongly reduced and shifted to higher temperature with increasing
applied magnetic field. This is the CMR effect. This type of behaviour is
seen in figure~\ref{fig:rho_T_g0.16} for $g/W=0.16$. The main differences
between theory and experiment are a more rapid observed drop in $\rho$ with
decreasing temperature below $\Tc$ and a more sensitive observed CMR effect.
$h/W=0.004$ corresponds to a field of about $20\,{\rm T}$ for $W=1\,{\rm eV}$
and the corresponding reduction in $\rho$ in figure~\ref{fig:rho_T_g0.16} is
achieved with a field of about $5\,{\rm T}$ experimentally. Millis \etal
\cite{MiMuSh96II} noted a similar problem in their work using classical
phonons. Both of the discrepancies mentioned might be remedied by introducing
a dependence of $g$ on $\rho$, corresponding to more efficient screening of
the electron-phonon interaction with increasing metallization. The huge
reduction in resistivity peak on reducing from $g/W$ from $0.16$ to $0.10$
shows the extreme sensitivity of $\rho$ to changes in $g$. The main point to
notice is that we can understand the enormous difference between \chem{LCMO}
and \chem{LSMO} within our theory by assuming the electron-phonon coupling
changes from $g/W=0.16$ in \chem{LCMO} to $g/W=0.10$, or slightly greater, in
\chem{LSMO}. The observed ratio of the Curie temperatures, slightly less than
$2$, is then in accord with figure~\ref{fig:Tc_coupling}. As discussed in
section \ref{sec:many-body-cpa-hde} the critical coupling $g_{\rm c}$ for the
formation of a polaron band is $g_{\rm c}/W\approx 0.15$, with our phonon
energy $\om/W=0.05$, and to obtain the right order of magnitude for $\rho$
above $\Tc$ in \chem{LCMO}, g is pinned down closely to $0.16$. A larger value
for $g$ leads to too high a resistivity and too low a Curie temperature. It
is interesting that neither $\rho(T)$ nor $\Tc$ change when $g/W$ is varied
between $0.1$ and $0$. This means that \chem{LSMO} can be described very well
by the pure DE model, as stressed by Furukawa \cite{phys_of_mns}, but it does
not follow that electron-phonon coupling is negligible. However, from the
results of Millis \etal \cite{MiMuSh96II} for classical phonons, one can
understand why a coupling small enough to give a \chem{LSMO}-like
$\rho(T)$ curve does not lead to a change in slope of the rms oxygen
displacements, as a function of temperature, at $\Tc$. No such change is
found in \chem{LSMO} \cite{MaEn96}, in contrast to the case of \chem{LCMO}
\cite{DaZhMo96}. It is more difficult to understand the observation
\cite{LoEgBr97} of static local Jahn-Teller distortions in \chem{LSMO} at
room temperature, apparently associated with localized carriers in the
presence of metallic conduction.

From figures~\ref{fig:DOS_g}
and \ref{fig:DOSs_PM_FM} we see that for $g/W=0.1$, appropriate to
\chem{LSMO}, there is no sign of a pseudogap in the density of states. An
actual gap in the hypothetical paramagnetic state at $T=0$ appears at
$g=g_{\rm c}$ with $g_{\rm c}$ between $0.15$ and $0.16$. From
figure~\ref{fig:DOSs_PM_FM} we see that for $g/W=0.16$, appropriate to
\chem{LCMO}, a few polaron subbands have appeared in the gap. These are seen
much more clearly in figure~\ref{fig:DOS_polarons} for $g/W=0.18$ when there
is a larger gap. However the subband structure is washed out completely for
$\beta W=25$, corresponding to $T=464\,{\rm K}$ for $W=1\,{\rm eV}$, and this
effect will occur at a much lower temperature for $g/W=0.16$. Thus in the
actual paramagnetic state of \chem{LCMO} above $\Tc$ we do not expect the
quantum nature of phonons to manifest itself, so we are essentially in the
classical regime of Millis \etal \cite{MiMuSh96II}. The same is true in the
saturated ferromagnetic state at $T=0$ where only a pseudogap appears in
figure~\ref{fig:DOSs_PM_FM}. As the temperature rises towards $\Tc$ a
minority spin band grows, also with a pseudogap, while the majority spin band
loses weight. The width of the ferromagnetic bands decreases with increasing
temperature, corresponding to the DE effect, but the narrowing in the
paramagnetic state is not so marked as in the pure DE model. Thus double
exchange is not so effective in the presence of strong electron-phonon
coupling, which is consistent with the reduction in $\Tc$ shown in
figure~\ref{fig:Tc_coupling}.

Clearly our picture of the manganites is close in spirit to that of Millis
\etal \cite{MiMuSh96II}, although the relationship to polaron physics is not
so clear in their classical approximation. Other authors adopt completely
different viewpoints. Furukawa \cite{phys_of_mns} rejects the importance of
electron-phonon coupling and argues in favour of phase separation models of
\chem{LCMO} with, for example, ferromagnetic and charge-ordered regions. Such
scenarios have been extensively discussed by the Florida group
\cite{MoreoYuDa99,MaMoVe00}. It is controversial whether such phase
separation can occur for $x\approx0.33$ which is far from the
antiferromagnetic insulating phase near $x=0$ and the region with charge and
orbital ordering near $x=0.5$. Nagaev \cite{Naga99} also argues against any
polaronic effects, while Alexandrov and Bratkowsky \cite{AlBr99,AlexBr99}
assume strong electron-phonon coupling with small polarons even in the
ferromagnetic state and with immobile bipolarons forming near $\Tc$. The
public correspondence \cite{Comment49,AlexBr99} between Nagaev and Alexandrov
and Bratkowsky (AB) centres on estimating the magnitude of the polaron
binding energy $E_{\rm p}$ and the criterion for small polaron formation
\cite{AlBrat99}. Since our picture of \chem{LCMO} lies between their extreme
views it is interesting to compare our estimates with theirs. For \chem{LCMO}
we find $E_{\rm p}=g^2/\om\approx0.5\,{\rm eV}$ for $W=1\,{\rm eV}$ whereas,
for manganites in general, Nagaev estimates $E_{\rm p}\approx0.1-0.3\,{\rm
  eV}$ and AB estimate $E_{\rm p}\approx1\,{\rm eV}$. Our condition for
small-polaron formation in a paramagnetic state at $T=0$ is $g>g_{\rm
  c}\approx0.15W$ which corresponds to $E_{\rm p}>0.45W$. Nagaev adopts the
criterion $E_{\rm p}>W$, remembering that $W$ is the half-bandwidth in our
notation, while AB \cite{AlBrat99} propose $E_{\rm p}>2W(8z)^{-1/2}=0.29W$
with number of nearest neighbours $z$ taken as $6$. AB's condition is less
stringent than Eagle's \cite{Eagl66} condition for ``nearly small polarons''
$E_{\rm p}>Wz^{-1/2}=0.41W$ which is close to ours. Both as regards this
criterion and the value of $E_{\rm p}$ for \chem{LCMO}, our results are
intermediate between Nagaev's and AB's, as expected. For \chem{LSMO}, on the
other hand, our estimate of $E_{\rm p}$ is $0.2\,{\rm eV}$. In this case we
agree with Nagaev that electron-phonon coupling is not important.
AB assume stronger electron-phonon coupling than we find in \chem{LCMO}.
Hence they have a coherent small-polaron band in the ferromagnetic state
which apparently supplies the required metallic conductivity. The carriers
are unable to bind to form singlet bipolarons.

AB \cite{AlBr99,AlexBr99} assume that the strong Hund's rule coupling, which
we and most authors assume is effectively infinite, can be treated in mean
field theory. Instead of the double-exchange effect, in which the occupied
polaron band merely narrows as its weight redistributes between up and down
spin, the Hund's rule exchange splitting collapses above $\Tc$ and it is
assumed that immobile singlet bipolarons can form. Hence near $\Tc$ there is
an enormous increase in resistivity which decreases above $\Tc$ as the
bipolarons dissociate. The problem with this theory as it stands, even more
serious than doubts about the existence of polarons above and below $\Tc$, is
the mean field treatment of the large Hund's rule exchange.

\begin{figure}
  \centering \includegraphics[width=0.75\textwidth]{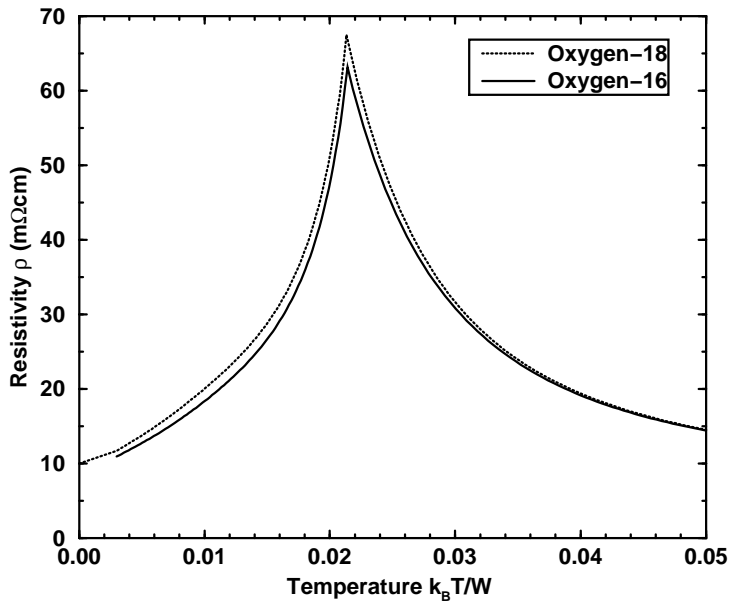}
    \caption{\label{fig:isotope_LCMO}%
      The absence of an isotope effect in \chem{LCMO} due to a simple mass
      scaling $\left(\propto M^{1/2}\right)$ of the phonon frequency.}
\end{figure}

\begin{figure}
  \centering \includegraphics[width=0.75\textwidth]{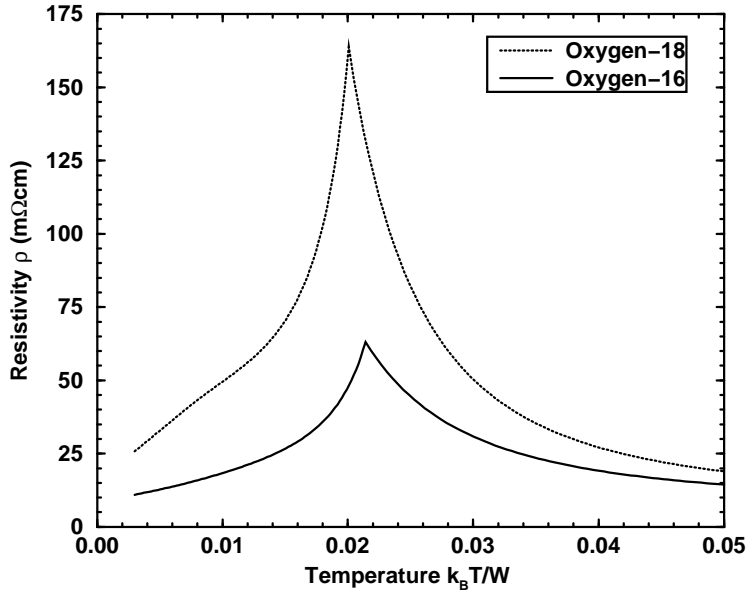}
    \caption{\label{fig:isotope_oxygen}%
      A giant isotope effect due to mass scaling of the phonon frequency
      together with a $3\%$ increase of electron-phonon coupling strength in
      the \chem{^{18}O} system. This arbitrary modelling shows how an effect
      similar to that observed \cite{ZhKaPr00} might arise from small
      parameter changes associated with changed oxygen content.}
\end{figure}

The giant isotope effect observed in \chem{LCMO} and
\chem{Nd_{0.7}Sr_{0.3}MnO_{3}} has been interpreted as evidence for immobile
bipolaron formation \cite{ZhKaPr00}. The effect is seen as a decrease of
$\Tc$ and a big increase in resistivity, particularly near $\Tc$, when
\chem{^{16}O} is replaced by \chem{^{18}O}. To investigate this effect in our
model some care is needed. It was pointed out that in
equation~(\ref{eq:h_hde}) the electron-phonon coupling term corresponds to a
term $-g'\sum_i n_i x_i$ where $x_i$ is to be associated with oxygen
displacement around a \chem{Mn} atom. Here $g'$ should be independent of the
oxygen mass $M$. In the second-quantized form of equation~(\ref{eq:h_hde})
one finds $g=g'\left(2M\om\right)^{-1/2}$ so that the polaron binding energy
$g^2/\om={g'}^2/\left(2M\om^2\right)$. Since for an oscillator $\om\propto
M^{-1/2}$ the polaron binding energy $g^2/\om$ is expected to be independent
of $M$.  However $g$ varies as $M^{-1/4}$. We have therefore recalculated the
resistivity $\rho(T)$ with the same parameters as used in
figure~\ref{fig:rho_T_g0.16} for \chem{^{16}O} but with $g$ and $\om$ scaled
appropriately for \chem{^{18}O}. The results are compared in
figure~\ref{fig:isotope_LCMO}. The almost complete absence of an isotope
effect in both $\Tc$ and $\rho$ is remarkable in view of the moderately
strong electron-phonon coupling in our model of \chem{LCMO}. Since we do not
believe in the bipolaron theory for \chem{LCMO} another explanation of the
isotope effect must be sought. Nagaev has considered several possibilities,
all based on the isotope dependence of the number of excess or deficient
oxygen atoms in thermodynamic equilibrium \cite{Naga99,Naga98}. In
\chem{LCMO} with $x=0.2$ there is experimental evidence that at least part of
the isotope effect is due to varying oxygen content \cite{FrIsCh98}. A change
in oxygen nonstoichiometry could lead to a change in carrier density and,
perhaps due to volume change, to a change in electron-phonon coupling $g'$.
In view of the sensitivity of $\rho(T)$ to electron-phonon coupling we think
that the isotope effect may be due more to a change in $g'$, or to a change in
$\om$ beyond the simple mass scaling, than to a change in $n$. Since Nagaev
believes electron-phonon coupling to be unimportant he did not consider this
particular consequence of his nonstoichiometry proposal. The involvement of
electron-phonon coupling in the isotope effect would explain why the $\Tc$
shift in \chem{LSMO} is much smaller than in \chem{LCMO} \cite{ZhCoKe96}. To
illustrate such a mechanism we have recalculated $\rho(T)$ for the
\chem{^{18}O} system with scaled $\om$ as before $\left(\propto
  M^{-1/2}\right)$ but with $g$ unchanged from its \chem{^{16}O} value. This
implies a $3\%$ increase in $g'$. The results are shown in
figure~\ref{fig:isotope_oxygen}. In the paramagnetic state they are quite
similar to those observed in \chem{LCMO} with $x=0.25$ and in
\chem{Nd_{0.7}Sr_{0.3}Mn_O3}. The only conclusion we can draw from this
arbitrary calculation is that it may be extremely difficult to determine the
true origin of the isotope effect in such a critical system as \chem{LCMO} on
the verge of small-polaron formation.  However it does not arise from a
simple mass scaling of the phonon frequency.

\begin{figure}
  \centering \includegraphics[width=0.75\textwidth]{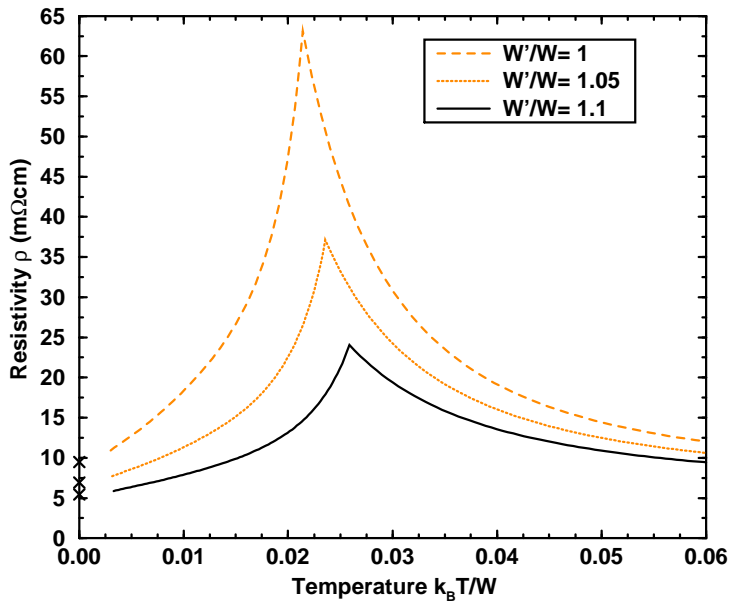}
    \caption{\label{fig:pressure}%
      The effect of pressure on the resistivity and Curie temperature in
      \chem{LCMO} by increasing the half-bandwidth $W$ to $W'=1.05W$ and
      $1.1W$.}
\end{figure}

Another example of the sensitivity of $\rho$ to variation in parameters is
the effect of pressure \cite{NeHuThHe95}. The strong suppression of the
resistivity peak and the increase in Curie temperature in \chem{LCMO} can be
modelled within our theory by increasing the bandwidth \cite{Gr01}, and
assuming other terms in the Hamiltonian are constant. Calculated results are
shown in figure~\ref{fig:pressure}. This pressure effect is due to a
reduction in $g^2/(\om W)$, the ratio of polaron binding energy to the
half-bandwidth. A simple estimate, using the known compressibility and
dependence of $W$ on lattice constant, shows that the theoretical pressure
for a given effect is about four times larger than that required
experimentally.  This is the same factor that we found in the case of the
magnetic field required for a given CMR effect, so both discrepancies could
possibly be removed with the same dependence of $g$ on $\rho$ , due to
screening, postulated earlier.

According to our theory we expect pseudogaps at the Fermi level to be
observable in the density of states of \chem{LCMO} both below and above
$\Tc$. These should appear in experiments such as scanning tunnelling
spectroscopy, photoemission and optical conductivity measurements. No pseudo
gaps are expected in \chem{LSMO}. The pseudogap is a feature of the atomic
limit, typified by the envelope function in figure~\ref{fig:hde_AL} for $T=0$
with maxima determined by the polaron binding energy $g^2/\om$, and is
completely washed out when $g^2/(\om W)<0.2$ (see figure~\ref{fig:DOS_g}).
Early results of scanning tunnelling spectroscopy on \chem{LCMO}
\cite{WeYeVa97} with $x=0.3$ seem unlikely to relate to the bulk. In the
ferromagnetic state at $77\,{\rm K}$ there is a huge gap of about $1\,{\rm
  eV}$. It is not clear why the authors interpret this as evidence for
half-metallic ferromagnetism. A gap of this size associated with
small-polaron formation in the bulk would imply an unrealistically large
electron-phonon coupling, certainly incompatible with metallic conduction and
a Curie temperature of reasonable magnitude for \chem{LCMO}. More recently
Biswas \etal \cite{BiElRaBh99} reported a scanning tunnelling spectroscopy
study of several manganites. The results are very much in accord with our
theory. There is no gap in the low temperature ferromagnetic state but a
small gap (pseudogap) appears for $T\approx\Tc$ in the low (high) $\Tc$
materials. As $T$ increases above $\Tc$ the pseudogap or gap gets filled in
as we would expect (see figure~\ref{fig:DOS_polarons}).

In an extremely interesting paper on angle-resolved photoemission
spectroscopy for the bilayer manganite \chem{La_{1.2}Sr_{1.8}Mn_2 O_7},
nominally with $n=0.6$, Dessau \etal \cite{De98} interpret their results very
much in the spirit of our theory. In this layered structure we cannot
estimate the strength of the electron-phonon coupling from $\Tc$, whose low
value of $126\,{\rm K}$ is largely a result of quasi-two dimensional
fluctuations. However the large low-temperature resistivity of more than
$3\,{\rm m\Omega\,cm}$ suggests that small-polaron bands might exist in both the
ferromagnetic and paramagnetic state. This indicates a larger value of $g/W$
than in the cubic manganites, which might be expected from a reduced
bandwidth in the layered structure. In this case the polaron bands would not
be washed out above $\Tc$ and should exist within a wider gap than in the
ferromagnetic state. This is compatible with the observations. Dessau \etal
\cite{De98} interpret the widths of their $\bi{k}$-resolved spectral peaks in
terms of multiple polaron subbands. The fact that strong peaks all occur at
$0.65-1.0\,{\rm eV}$ below the Fermi level, independent of $\bi{k}$, suggests
that the system is not so far from the atomic limit, with the spectral peaks
given by the envelope function in figure~\ref{fig:hde_AL} for $T=0$. This
indicates a large polaron binding energy of $0.65-1.0\,{\rm eV}$ and
Alexandrov and Bratkovsky have applied their theory \cite{AlBrat99} to these
bilayer manganites. They interpret the observed upward shift in frequency
$\nu$ of the maximum in the optical conductivity $\si(\nu)$, between
$10\,{\rm K}$ and just above $\Tc$, as due to bipolaron formation. We cannot
discuss the possible binding of two polarons on nearest-neighbour sites
within our local approximation. However one does not need bipolarons to
understand an upward shift of the optical conductivity maximum with
increasing temperature. This is clearly seen already in the calculations of
Millis \etal \cite{MiMuSh96II} where the phonons are treated classically. In
connection with the work of Dessau \etal \cite{De98}, it should be mentioned
that Moreo \etal \cite{MoYu99} interpret the observed pseudogap not as an
intrinsic property but in terms of phase separation.

Kim \etal \cite{KiJu98} have reported very interesting measurements of
optical conductivity $\si(\nu)$ in \chem{La_{0.7-y}Pr_y Ca_{0.3}MnO_3} for
$y=0$, $0.13$, $0.4$ and $0.5$. The Curie temperature takes corresponding
values of approximately $245$, $240$, $155$ and $120\,{\rm K}$. With our
parameterization this corresponds, from figure~\ref{fig:Tc_coupling}, to
$g/W$ varying between $0.16$ and $0.22$. The observed variation of $\si(\nu)$
with $y$ and temperature $T$, over a wide range of photon energy up to
$2\,{\rm eV}$, is in general agreement with the classical treatment of Millis
\etal \cite{MiMuSh96II}. It is possible that the unusual behaviour in the far
infrared region below $0.15\,{\rm eV}$, for $y=0.4$ and $0.5$, may relate to
our calculated electronic structure in the pseudogap in addition to direct
excitation of phonons. Detailed calculations of angle-resolved photoemission
and optical conductivity spectra, and their comparison with experimental
data, will be reported later.


\section{Conclusion}\label{sec:conclusion}

We have summarized the many-body CPA treatment of the double-exchange model,
with and without coupling to local phonons, which is presented in detail in
three earlier papers \cite{EdGrKu99,Gr01,GrEd99}.  The method can deal with
quantum local spins and phonons and is equivalent to dynamical mean field
theory in the classical limit. For quite a wide range of fairly weak
electron-phonon coupling we find that the results are essentially those of
the pure double-exchange model. As the electron-phonon coupling is increased
our theory can describe, for the first time, the crossover from the classical
phonon limit \cite{MiMuSh96II} to the formation of small-polaron bands, and
finally it links up with standard small-polaron theory in the strong-coupling
limit. In section \ref{sec:appl-mang} we have given a much fuller discussion
of the application to manganites than in the earlier papers. As well as basic
properties such as Curie temperature, resistivity and magnetoresistance, we
discuss the interpretation of the giant isotope effect, pressure effects and
pseudogaps observed in scanning tunnelling spectroscopy, photoemission and
optical conductivity.

We find that a typical manganite like \chem{La_{1-x}Ca_x MnO_3}, with
$x\approx0.33$, is in the critical regime on the verge of small-polaron
formation. This explains its extreme sensitivity to changes in parameters
arising from isotopic substitution and pressure. On the other hand
\chem{La_{1-x}Sr_xMnO_3}, with weaker electron-phonon coupling, is quite well
described by the pure double-exchange model, as pointed out by Furukawa
\cite{phys_of_mns}.  The large pseudogap seen in a bilayer manganite
\cite{De98} indicates stronger electron-phonon coupling than in most cubic
manganites.

We are grateful to Lesley Cohen, D.M. Eagles, K. Kamenev, N. Furukawa and
K. Kubo for helpful discussion and to the UK Engineering and Physical
Sciences Research Council (EPSRC) for financial support. Elizabeth Rowley has
given valuable help in preparing the manuscript for publication.




\end{document}